\documentclass[jou,10pt]{apa6}

\usepackage{amsmath}
\usepackage{amsfonts}
\usepackage{amssymb}
\usepackage{listings}
\makeatletter
\def\Let@{\def\\{\notag\math@cr}}
\makeatother
\usepackage{graphicx}
\usepackage[utf8]{inputenc}
\usepackage[T1]{fontenc}
\usepackage{float}

\usepackage[utf8]{inputenc}
\usepackage{amsmath,amssymb}
\newcommand{\textVerb}[1]{\texttt{\mbox{#1}}}
\usepackage{subcaption}
\usepackage{algorithm,algorithmic}
\usepackage{color}

\usepackage[american]{babel}
\usepackage{csquotes}

\usepackage{natbib}
\usepackage{caption}
\usepackage{{url}}
\usepackage{rotating}

\title{A Tutorial on Regularized Partial Correlation Networks}

\shorttitle{REGULARIZED PARTIAL CORRELATION NETWORKS}

\author{Sacha Epskamp and Eiko I.\ Fried}

\affiliation{University of Amsterdam: Department of Psychological Methods}

\usepackage{tocloft}
\makeatletter
\renewcommand{\@cftmaketoctitle}{}
\makeatother

\abstract{
\begin{center}
\tableofcontents
\end{center}
}

\begin{document}

\maketitle

\raggedbottom
\urlstyle{same}

\section{Abstract}

Recent years have seen an emergence of network modeling applied to moods, attitudes, and problems in the realm of psychology. In this framework, psychological variables are understood to directly affect each other rather than being caused by an unobserved latent entity. In this tutorial, we introduce the reader to estimating the most popular network model for psychological data: the partial correlation network. We describe how regularization techniques can be used to efficiently estimate a parsimonious and interpretable network structure in psychological data. We show how to perform these analyses in R and demonstrate the method in an empirical example on post-traumatic stress disorder data. In addition, we discuss the effect of the hyperparameter that needs to be manually set by the researcher, how to handle non-normal data, how to determine the required sample size for a network analysis, and provide a checklist with potential solutions for problems that can arise when estimating regularized partial correlation networks.

\section{Introduction}

Recent years have seen increasing use of network modeling for exploratory studies of psychological behavior as an alternative to latent variable modeling \citep{borsboom2013network,schmittmann2013}. In these so-called \emph{psychological networks} \citep{bootnetpaper}, nodes represent psychological variables such as mood states, symptoms, or attitudes, while edges (links connecting two nodes) represent unknown statistical relationships that need to be estimated. As a result, this class of network models is strikingly different from social networks, in which edges are known \citep{wasserman1994}, posing novel problems for statistical inference. A great body of technical literature exists on the estimation of network models (e.g., \citealt{meinshausen2006high,friedman2008sparse,hastie01statisticallearning,hastie2015statistical,foygel2010extended}). However, this line of literature often requires a more technical background and does not focus on the unique problems that come with analyzing psychological data, such as the handling of ordinal data, how a limited sample size affects the results, and the correspondence between network models and latent variable models.

Currently, the most common model used to estimate psychological networks based on continuous data is the \emph{partial correlation network}. Partial correlation networks are usually estimated using \emph{regularization} techniques originating from the field of machine learning. These techniques have been shown to perform well in retrieving the true network structure \citep{meinshausen2006high,friedman2008sparse,foygel2010extended}. Regularization involves estimating a statistical model with an extra penalty for model complexity. Doing so leads to a model to be estimated that is \emph{sparse}: many parameters are estimated to be exactly zero. When estimating networks, this means that edges that are likely to be spurious are removed from the model, leading to networks that are  simpler to interpret. Regularization therefore jointly performs model-selection and parameter estimation. Regularization techniques have grown prominent in many analytic methods, ranging from regression analysis to principal component analysis \citep{hastie2015statistical}. In this tutorial, we will only discuss regularization in the context of network estimation. For an overview of such methods applied more broadly in psychological methods, we refer the reader to \citet{chapman2016statistical}.

  Regularized network estimation has already been used in a substantive number of publications in diverse fields, such as clinical psychology (e.g., \citealt{fried2016,knefel2016association,levine2016identifying,jaya2015loneliness,Deserno17082016,van2015association,Deserno17082016}), psychiatry (e.g., \citealt{isvoranu,isvoranua,mcnally2016can}), personality research (e.g., \citealt{costantini2015state,costantini2015development}), and health sciences (e.g., \citealt{kossakowski2015, langley2015should}). What these papers have in common is that they assume observed variables to causally influence one another, leading to network models consisting of nodes such as psychopathology symptoms (e.g. sad mood, fatigue and insomnia), items of personality domains like conscientiousness (e.g. impulse-control, orderliness and industriousness), or health behaviors (e.g. feeling full of energy, getting sick easily, having difficulties performing daily tasks).
From this network perspective, correlations among items stem from mutual associations among variables, which differs from the traditional perspective where latent variables are thought to explain the correlation among variables \citep{schmittmann2013}. Psychological networks thus offer a different view of item clusters: syndromes such as depression or anxiety disorder in the realm of mental disorders \citep{cramer2010comorbidity,borsboom2017network,friedreview}, personality facets or domains such as extraversion or neuroticism in personality research \citep{mottus2017traits,cramer2012dimensions}, health domains like physical or social functioning in health research \citep{kossakowski2015}, and the $g$-factor in intelligence research \citep{van2006dynamical,van2017network}. Important to note is that one does not have to adhere to this network perspective (i.e., network theory) in order to use the methods described in this tutorial (i.e. network methodology): psychological networks can be powerful tools to explore multicollinearity and predictive mediation, and can even be used to highlight the presence of latent variables. 

We are not aware of concise and clear introductions aimed at empirical researchers that explain regularization. The goal of this paper is thus (1) to provide an introduction to regularized partial correlation networks, (2) to outline the commands used in R to estimate these models, and (3) to address the most common problems and questions arising from analyzing regularized partial correlation networks.
%, such as power analysis and how to deal with non-normal or ordered-categorical data. 
The methodology introduced in this tutorial comes with the assumption that the cases (the rows of the spreadsheet) in the data set are independent, which is usually the case in cross-sectional data. Applying these methods to time-series data does not take temporal dependence between consecutive cases into account. We refer the reader to \citet{dynamics} to a discussion of extending this framework to such temporally ordered datasets. While this tutorial is primarily aimed at empirical researchers in psychology, the methodology can readily be applied to other fields of research as well.

This tutorial builds on the work of two prior tutorials: 
\citet{costantini2015state} focused on psychological networks in the domain of personality research, described different types of networks ranging from correlation networks to adaptive LASSO networks \citep{zou2006adaptive,parcor}, 
and introduced basic concepts such as centrality estimation in R. \citet{bootnetpaper} introduced several tests that allow researchers to investigate the accuracy and stability of psychological networks and derived graph-theoretical measures such as centrality, tackling the topics of generalizability and replicability. The present tutorial goes beyond these papers in the following ways: 
\begin{itemize}
\item \citet{costantini2015state} estimated the network structure using a different form of regularization (adaptive LASSO; \citealt{zou2006adaptive}), a different method for estimating the parameters (node-wise regressions; \citealt{meinshausen2006high}), and a different method for selecting the regularization tuning parameter (cross-validation; \citealt{parcor}). While an acceptable method for estimating regularized partial correlation networks, this procedure can lead to unstable results due to differences in the cross-validation sample selection (see section 2.5.6 of \citealt{costantini2015state}) and is not capable of handling ordinal data. We estimate regularized partial correlation networks via the EBIC graphical LASSO \citep{foygel2010extended}, using polychoric correlations as input when data are ordinal. We detail advantages of this methodology, an important one being that it can be used with ordinal variables that are very common in psychological research.
\item We offer a detailed explanation of partial correlations and how these should be interpreted. Especially since the work of \citet{costantini2015state}, researchers have gained a better understanding of the interpretation of partial correlation networks and their correspondence to multicollinearity and latent variable modeling. We summarize the most recent insights in these topics.
\item We provide a state-of-the-art FAQ addressing issues that researchers regularly struggle with---including power analysis and sample size recommendations that have been called for repeatedly \citep{bootnetpaper,friedchallenge}---and offer novel solutions to these challenges. 
\end{itemize}

The following sections are structured as follows. First, we introduce partial correlation networks and their estimation, providing detailed information on how these networks can be interpreted. Second, we explain regularization, an integral step in the estimation of partial correlation networks to avoid spurious relationships among items. Third, we explain how to best deal with non-normal (e.g., ordinal) data when estimating partial correlation networks. Fourth, we show researchers how to estimate partial correlation networks in R using an empirical example dataset. The fifth section covers replicability and power analysis for partial correlation networks. In this section, we present the simulation tool \verb|netSimulator|, which allows researchers to determine the sample size that would be required to successfully examine a specific network structure. We also summarize post-hoc stability and accuracy analyses that are described in detail elsewhere \citep{bootnetpaper}. Finally, we conclude with solutions to the most commonly encountered problems when estimating network models and cover supplemental topics such as comparing networks given unequal sample sizes, unexpected negative relationships among items, very strong positive relationships, or empty networks without any edges.

\section{Partial Correlation Networks}

The most commonly used framework for constructing a psychological network on data that can be assumed to be multivariate normal is to estimate a network of \emph{partial correlation coefficients} \citep{mcnally2015mental,borsboom2013network}. These coefficients range from $-1$ to $1$ and encode the remaining association between two nodes after controlling for all other information possible, also known as conditional independence associations. Partial correlation networks have also been called \emph{concentration graphs} \citep{cox1994note} or \emph{Gaussian graphical models} \citep{lauritzen1996graphical}, and are part of a more general class of statistical models termed \emph{pairwise Markov random fields} (see e.g., \citealt{koller2009probabilistic} and \citealt{murphy2012machine} for an extensive description of pairwise Markov random fields). The interpretation of partial correlation networks has recently been described in the psychological literature (e.g., conceptual guides are included in \citealt{costantini2015state} and the online supplementary materials of \citealt{bootnetpaper}; an extensive technical introduction is included in \citealt{dynamics}). To keep the present tutorial self-contained, we succinctly summarize the interpretation of partial correlations below. 

% Each edge in the partial correlation network represents a partial correlation coefficient between two variables after conditioning on all other variables in the dataset. These coefficients range from $-1$ to $1$ and encode the remaining association between two nodes after controlling for all other information possible, also known as conditional independence associations. Typically, the edges are visualized using red lines indicating negative partial correlations, green (or blue) lines indicating positive partial correlations, with wider and more saturated lines indicating stronger partial correlations \citep{jssv048i04}. Whenever the partial correlation is exactly zero, no edge is drawn between two nodes, indicating that two variables are independent after controlling for all other variables in the network.

%%%%

\paragraph{Drawing partial correlation networks} After partial correlations have been estimated, they can be visualized in a weighted network structure. Each node represents a variable and each edge represents that two variables are not independent after conditioning on all variables in the dataset. These edges have a weight, \emph{edge weights}, which are the partial correlation coefficients described below. Whenever the partial correlation is exactly zero, no edge is drawn between two nodes, indicating that two variables are independent after controlling for all other variables in the network. Several different software packages can be used to visualize the network. For example, one could use the freely available software packages \emph{cytoscape} \citep{shannon2003cytoscape}, \emph{gephi} \citep{bastian2009gephi}, or R packages \emph{qgraph} \citep{jssv048i04}, \emph{igraph} \citep{igraph}, \emph{Rgraphviz} \citep{rgraphviz} or \emph{ggraph} \citep{ggraph}. The \emph{qgraph} package has commonly been used in psychological literature as it automates many steps for drawing weighted networks and includes the estimation methods discussed in this paper. When drawing a network model, the color and weight of an edges indicates its magnitude and direction. Using \emph{qgraph}, red lines indicate negative partial correlations, green (using the classic theme) or blue (using the colorblind theme) lines indicate positive partial correlations, with wider and more saturated lines indicating stronger partial correlations \citep{jssv048i04}.

%%%%

\paragraph{Obtaining partial correlation networks} While multiple ways exist to compute partial correlation coefficients \citep{cohen2013applied}, we focus on two commonly-used methods that have been shown to obtain the partial correlations quickly. First, the partial correlations can be directly obtained from the inverse of a variance--covariance matrix. Let $\pmb{y}$ represent a set of item responses, which we can assume without loss of generality to be centered. Let $\pmb{\Sigma}$ (sigma) denote a variance--covariance matrix. Then, the following states that we assume $\pmb{y}$ to have a multivariate normal distribution:
\[
\pmb{y} \sim N( \pmb{0}, \pmb{\Sigma}).
\]
Let $\pmb{K}$ (kappa) denote the inverse of $\pmb{\Sigma}$, also termed the \emph{precision matrix}:
\[
\pmb{K} = \pmb{\Sigma}^{-1},
\]
then, element $\kappa_{ij}$ (row $i$, column $j$ of $\pmb{K}$) can be standardized to obtain the partial correlation coefficient between variable $y_i$ and variable $y_j$, after conditioning on all other variables in $\pmb{y}$, $\pmb{y}_{-(i,j)}$ \citep{lauritzen1996graphical}:
\[
\mathrm{Cor}\left(y_i, y_j \mid \pmb{y}_{-(i,j)}\right) = - \frac{\kappa_{ij}}{\sqrt{\kappa_{ii}}\sqrt{\kappa_{jj}}}.
\]
An alternative way to obtain the partial correlation coefficients is by using \emph{node-wise regressions} \citep{meinshausen2006high}. If one was to perform a multiple regression in which $y_1$ is predicted from all other variables:
\[
y_1 = \beta_{10} + \beta_{12} y_2 + \beta_{13} y_3 + \dots  + \varepsilon_1,
\]
followed by a similar regression model for $y_2$:
\[
y_2 = \beta_{20} + \beta_{21} y_1 + \beta_{23} y_3 + \dots  + \varepsilon_2,
\]
and similarly for $y_3$, $y_4$, etcetera, then, the same partial correlation coefficient between $y_i$ and $y_j$ is proportional to either the regression slope predicting $y_i$ from $y_j$ or the regression slope predicting $y_j$ from $y_i$ \citep{pourahmadi2011covariance}:
\[
\mathrm{Cor}\left(y_i, y_j \mid \pmb{y}_{-(i,j)}\right) = \frac{\beta_{ij} \mathrm{SD}(\varepsilon_j)}{\mathrm{SD}(\varepsilon_i)} = \frac{\beta_{ji} \mathrm{SD}(\varepsilon_i)}{\mathrm{SD}(\varepsilon_j)},
\]
in which SD stands for the standard-deviation. Obtaining a partial correlation coefficient by standardizing the precision matrix or performing node-wise regressions will lead to the exact same estimate. 

\paragraph{Interpreting partial correlation networks} Partial correlation networks allow for several powerful inferences. These points are a summary of a more detailed and technical introduction by \citet{dynamics}:
\begin{itemize}
\item \emph{Partial correlation networks allow one to model unique interactions between variables.}  
If $A$ correlates with $B$, and $B$ correlates with $C$, then we would naturally expect $A$ to correlate with $C$. An unconditional correlation of zero between $A$ and $C$ would be unexpected as only few causal structures would lead to such a correlational pattern.\footnote{Two possible options are if $B$ is a common effect of $A$ and $C$ or if two orthogonal latent variables cause covariation between $A$ and $B$ and between $B$ and $C$.} If the data are normal, partial correlations can be interpreted as \emph{pairwise interactions}\footnote{Not to be confused with interaction effects of two variables on an outcome variable.}, of which we only need two to model the correlational pattern: an interaction between $A$ and $B$ and an interaction between $B$ and $C$. This model will contain one degree of freedom and thus leads to a testable hypothesis \citep{epskampPsychometrika}. Such a point of view is akin to loglinear modeling of categorical data \citep{agresti2014categorical, wickens2014multiway}, which is structurally comparable to the partial correlation network \citep{netpsych}. 
\item \emph{The partial correlation network maps out multicollinearity and predictive mediation.} As shown above,  partial correlations are closely related to coefficients obtained in multiple regression models: when an independent variable does not predict the dependent variable, we would not expect an edge in the network. The strength of the partial correlation is furthermore directly related to the strength of the regression coefficient. The edges connected to a single node therefore show the researcher the expected result of a multiple regression analysis. Unlike what can be seen from a multiple regression analysis of a single dependent variable, however, the network also shows which variables would predict the independent variables. By linking separate multiple regression models, partial correlation networks allow for mapping out linear prediction and multicollinearity among all variables. This allows for insight into predictive mediation: a network in which two variables are not directly connected but are indirectly connected (e.g., $A$ -- $B$ -- $C$) indicates that $A$ and $C$ may be correlated, but any predictive effect from $A$ to $C$ (or vice versa) is mediated by $B$.
\item \emph{Partial correlations can be indicative of potential causal pathways.} Conditional independence relationships, such as those encoded by partial correlation coefficients, play a crucial role in causal inference \citep{pearl2000causality}. When all relevant variables are assumed to be observed (i.e., no latent variables), a partial correlation between variables $A$ and $B$ would only be expected to be non-zero if $A$ causes $B$, $B$ causes $A$, there is a reciprocal relationship between $A$ and $B$, or both $A$ and $B$ cause a third variable in the network \citep{pearl2000causality,koller2009probabilistic}. To this end, partial correlation networks are thought of as highly exploratory hypothesis-generating structures, indicative of potential causal effects.
While exploratory algorithms exist that aim to discover directed (causal) networks, they rely on strong assumptions such as acyclity (a variable may not eventually cause itself (e.g., $A \rightarrow B \rightarrow C \rightarrow A$), and are more strongly influenced by latent variables causing covariation (latent variables would induce directed edges between observed variables implying a strong causal hypothesis). Additionally, these models are not easily identified or parameterized: many equivalent directed models can fit the data equally well, all differently parameterized. Partial correlation networks, on the other hand, are well identified (no equivalent models) and easily parameterized using partial correlation coefficients. As such, exploratively estimating undirected networks offer an attractive alternative to exploratively estimating directed networks, without the troublesome and poorly identified direction of effect\footnote{A partial correlation network should not be interpreted to equate the skeleton of a causal model (a directed network with arrowheads removed), as conditioning on a common effect can induce an edge in the partial correlation network.  In addition, latent variables can induce edges in both  directed and  undirected networks. We discuss both common effects and latent variables in detail below.}.
\item \emph{Clusters in the network may highlight latent variables.} While partial correlations aim to highlight unique variance between two variables, they retain shared variance due to outside sources that cannot fully be partialled out by controlling for other variables in the network. As a result, if a latent variable causes covariation between two or more variables in the network, it is expected that all these variables will be connected in the network, forming a cluster \citep{golino2017exploratory}. Such clusters can thus be indicative of latent variables \citep{dynamics}. We discuss the relationship between networks and latent variable models in more detail at the end of this paper.
\end{itemize}

\section{LASSO regularization}

\paragraph{Limiting spurious edges} As shown above, partial correlations can readily be estimated by inverting the sample variance--covariance matrix or by performing sequential multiple regressions and standardizing the obtained coefficients. Estimating parameters from data, however, always comes with sampling variation, leading to estimates that are never exactly zero. Even when two variables are conditionally independent, we still obtain nonzero (although typically small) partial correlations that will be represented as very weak edges in the network. These edges are called \emph{spurious} or \emph{false positives} \citep{costantini2015state}. In order to prevent over-interpretation and failures to replicate estimated network structures, an important goal in network estimation is to limit the number of spurious connections. One way to do so is to test all partial correlations for statistical significance and remove all edges that fail to reach significance \citep{drton2004model}. However, this poses a problem of multiple testing, and correcting for this problem (e.g., by using a Bonferroni correction) results in a loss of power \citep{costantini2015state}\footnote{Unregularized partial correlations can also be seen to already reduce spurious edges in a network comprised of marginal correlation coefficients \citep{costantini2015state}.}.

% \subsection{Controlling for Spurious Connections: The LASSO}

\paragraph{The LASSO} 
An increasingly popular method for limiting the number of spurious edges---as well as for obtaining more interpretable networks that better extrapolate to new samples---is to use statistical \emph{regularization} techniques. An especially prominent method of regularization is the `least absolute shrinkage and selection operator' (LASSO; \citealt{tibshirani1996regression}), which, unlike other regularization techniques, can lead to parameter estimates of exactly zero. In essence, the LASSO limits the sum of absolute partial correlation coefficients; as a result, all estimates shrink, and some become exactly zero. More technically, if $\pmb{S}$ represents the sample variance--covariance matrix, LASSO aims to estimate $\pmb{K}$ by maximizing the \emph{penalized} likelihood function \citep{friedman2008sparse}:
\[
\log \mathrm{det} \left( \pmb{K} \right) - \mathrm{trace}\left(\pmb{S}\pmb{K} \right) - \lambda \sum_{<i,j>} |\kappa_{ij}|
\]
Alternatively, LASSO regularization can be applied on the individual regression models if a network is estimated using node-wise regressions \citep{meinshausen2006high}\footnote{In regularized node-wise regressions, partial correlations obtained from the regression model for one node might slightly differ from partial correlations obtained from the regression model for another node. A single estimate can then be obtained by averaging the two estimated partial correlations.}. Using the LASSO results in a \emph{sparse} network in which likely spurious edges are removed \citep{boschlooCommentary}. The LASSO utilizes a tuning parameter $\lambda$ (lambda) that controls the level of sparsity. As can be seen above, $\lambda$ directly controls how much the likelihood is penalized for the sum of absolute parameter values. When the tuning parameter is low, only a few edges are removed, likely resulting in the retention of spurious edges. When the tuning parameter is high, many edges are removed, likely resulting in the removal of true edges in addition to the removal of spurious edges. The tuning parameter therefore needs to be carefully selected to create a network structure that minimizes the number of spurious edges while maximizing the number of true edges \citep{foygel2014high, foygel2010extended}.

%Typically, a range of networks is estimated under different values of $\lambda$ (Zhao \& Yu, \citeyear{zhao2006model}). The value for $\lambda$ under which no edges are retained (the empty network), $\lambda_{\mathrm{max}}$, is set to the largest absolute correlation (Zhao et al., \citeyear{huge}). Next, a minimum value can be chosen by multiplying some ratio $R$ with this maximum value\footnote{The \emph{qgraph} package uses $R = 0.01$ by default.}:
%\[
%\lambda_{\mathrm{min}} = R \lambda_{\mathrm{max}}.
%\]
%A logarithmically spaced range of tuning parameters (typically $100$ different values), ranging from $\lambda_{\mathrm{min}}$ to $\lambda_{\mathrm{max}}$, can be used to estimate different networks. To summarize, 

\paragraph{Selecting the LASSO tuning parameter} Typically, several networks are estimated under different values of $\lambda$ \citep{zhao2006model}. The different $\lambda$ values can be chosen from a logarithmically spaced range between a maximal $\lambda$ value for which no edge is retained (when $\lambda$ equals the largest absolute correlation; \citealt{huge}), and some scalar times this maximal $\lambda$ value\footnote{Current \emph{qgraph} package version 1.4.4 uses 0.01 as scalar and estimates 100 networks by default.}. Thus, the LASSO is commonly used to estimate a \emph{collection} of networks rather than a single network, ranging from a fully connected network to a fully disconnected network. Next, one needs to select the best network out of this collection of networks. This selection can be done by optimizing the fit of the network to the data by minimizing some information criterion. Minimizing the Extended Bayesian Information Criterion (EBIC; \citealt{chen2008EBIC}) has been shown to work particularly well in retrieving the true network structure \citep{foygel2014high, foygel2010extended, van2014new}, especially when the generating network is sparse (i.e., does not contain many edges). LASSO regularization with EBIC model selection has been shown to feature high specificity all-around (i.e., not estimating edges that are not in the true network) but a varying sensitivity (i.e., estimating edges that are in the true network) based on the true network structure and sample size. For example, sensitivity typically is less when the true network is dense (contains many edges) or features some nodes with many edges (hubs).

\paragraph{Choosing the EBIC hyperparameter} The EBIC uses a hyperparameter\footnote{A hyperparameter is a parameter that controls other parameters, and usually needs to be set manually.} $\gamma$ (gamma) that controls how much the EBIC prefers simpler models (fewer edges; \citealt{chen2008EBIC,foygel2010extended}):
\[
\mathrm{EBIC} = -2 L + E \log(N) + 4 \gamma E  \log(P),
\]
in which $L$ denotes the log-likelihood, $N$ the sample size, $E$ the number of non-zero edges and $P$ the number of nodes. This hyperparameter $\gamma$ should not be confused with the LASSO tuning parameter $\lambda$, and needs to be set manually. It typically is set between 0 and 0.5 \citep{foygel2010extended}, with higher values indicating that simpler models (more parsimonious models with fewer edges) are preferred. Setting the hyperparameter to $0$ errs on the side of discovery: more edges are estimated, including possible spurious ones (the network has a higher sensitivity). Setting the hyperparameter to $0.5$, as suggested by \citet{foygel2010extended}, errs on the side of caution or parsimony: fewer edges are obtained, avoiding most spurious edges but possibly missing some edges (i.e., the network has a higher specificity). Even when setting the hyperparameter to 0, the network will still be sparser compared to a partial correlation network that does not employ any form of regularization; setting $\gamma$ to 0 indicates that the EBIC reduces to the standard BIC, which still prefers simple models.

Many variants of the LASSO have been implemented in open-source software (e.g., \citealt{parcor, huge}). We suggest the variant termed the `graphical LASSO' (glasso; \citealt{friedman2008sparse}), which is specifically aimed at estimating partial correlation networks by inverting the sample variance--covariance matrix. The glasso algorithm has been implemented in the \emph{glasso} package \citep{glasso} for the statistical programming language R \citep{R}. A function that uses this package in combination with EBIC model selection as described by \citet{foygel2010extended} has been implemented in the R package \emph{qgraph} (Epskamp et al., 2012), and can be called via the \emph{bootnet} package \citep{bootnetpaper}. The glasso algorithm directly penalizes elements of the variance--covariance matrix, which differs from other LASSO network estimation methods which typically aim to estimate a network structure by penalizing regression coefficients in a series of multiple regression models \citep{meinshausen2006high}. We suggest using this routine because it can be engaged using simple input commands and because it only requires an estimate of the covariance matrix and not the raw data, allowing one to use polychoric correlation matrices when the data are ordinal (discussed below). 

\begin{figure}
\centering
\includegraphics[width=0.9\linewidth]{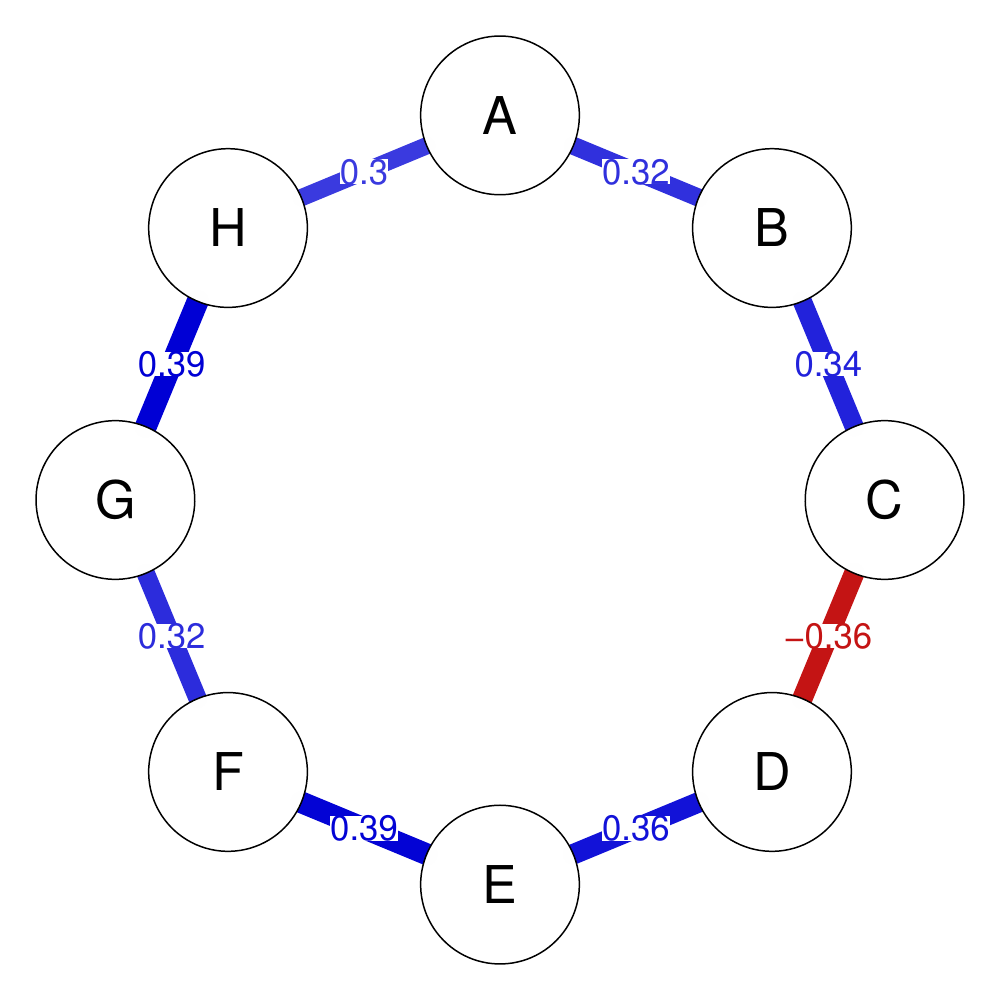}
\caption{True network structure used in simulation example. The network represents a \emph{partial correlation network}: nodes represent observed variables and edges represent partial correlations between two variables after conditioning on all other variables. The simulated structure is a \emph{chain graph} in which all absolute partial correlation coefficients were drawn randomly between $0.3$ and $0.4$.}
\label{primer:fig:1}
\end{figure}

\begin{figure*}
\centering
\includegraphics[width=1\textwidth]{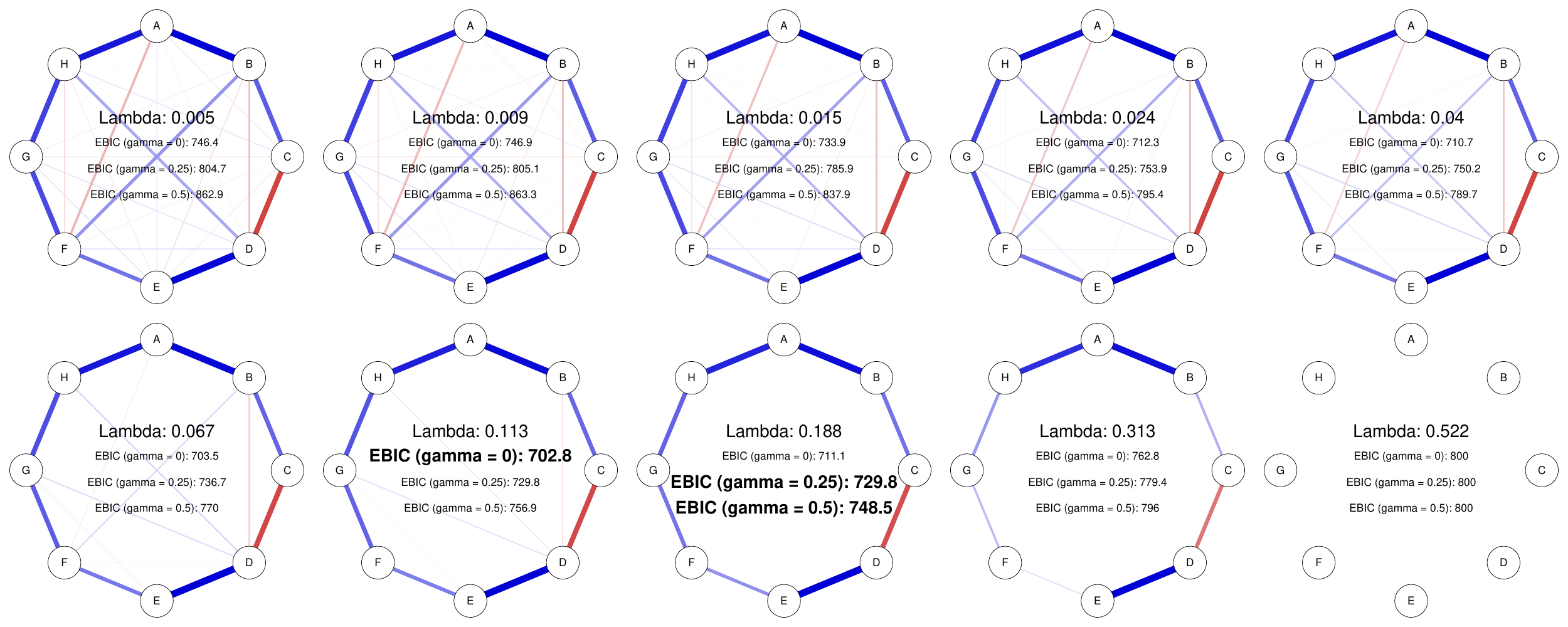}
\caption{Ten different partial correlation networks estimated using LASSO regularization. Setting the LASSO tuningparameter $\lambda$ that controls sparsity leads to networks ranging from densely connected to fully unconnected. Data were simulated under the network represented in Figure~\ref{primer:fig:1}. The fit of every network was assessed using the EBIC, using hyperparameter $\gamma$ set to $0$, $0.25$ or $0.5$. The bold-faced EBIC value is the best, indicating the network which would be selected and returned using that $\gamma$ value.}
\label{primer:fig:2}
\end{figure*}

To exemplify the above-described method of selecting a best-fitting regularized partial correlation network, we simulated a dataset of 100 people and 8 nodes (variables) based on the \emph{chain graph} shown in Figure~\ref{primer:fig:1}. Such graphs are particularly suitable for our example because the true network (the one we want to recover with our statistical analysis) only features edges among neighboring nodes visualized in a circle. This makes spurious edges---any edge that connects non-neighboring nodes---easy to identify visually. We used the \emph{qgraph} package to estimate 100 different network structures, based on different values for $\lambda$, and computed the EBIC under different values of $\gamma$. Figure~\ref{primer:fig:2} depicts a representative sample of 10 of these networks. Networks 1 through 7 feature spurious edges and err on the side of discovery, while networks 9 and 10 recover too few edges and err on the side of caution. For each network, we computed the EBIC based on $\gamma$ of 0, 0.25 and 0.5 (the hyperparameter the researchers needs to set manually). The boldface values show the best fitting models, indicating which models would be selected using a certain value of $\gamma$.  When $\gamma=0$ was used, network 7 was selected that featured three weak spurious edges. When $\gamma$ was set to 0.25 or 0.5 (the latter being the default in \emph{qgraph}) respectively, network 8 was selected, which has the same structure as the true network shown in Figure~\ref{primer:fig:1}. These results show that in our case, varying $\gamma$ changed the results only slightly. Importantly, this simulation does not imply that $\gamma=0.5$ always leads to the true model; simulation work has shown that 0.5 is fairly conservative and may result in omitting true edges from the network \citep{foygel2010extended}. In sum, the choice of the hyperparameter is somewhat arbitrary and up to the researcher, and depends on the relative importance assigned to caution or discovery \citep{dziak2012sensitivity}. Which of these $\gamma$ values work best is a complex function of the (usually unknown) true network structure.

\paragraph{A note on sparsity} It is important to note that although LASSO regularization\footnote{These arguments apply for other frequentist model selection methods as well, such as removing edges based on statistical significance.} will lead to edges being removed from the network, it does not present evidence that these edges are, in fact, zero \citep{boschlooCommentary}. This is because LASSO seeks to maximize \emph{specificity}; that is, it aims to include as few \emph{false positives} (edges that are not in the true model) as possible. As a result, observing an estimated network that is sparse (containing missing edges), or even observing an empty network, is in no way evidence that there are, in fact, missing edges. LASSO estimation may result in many \emph{false negatives}, edges that are not present in the estimated network but are present in the true network. This is related to a well-known problem of null hypothesis testing: \emph{Not} rejecting the null-hypothesis is not evidence that the null hypothesis is true \citep{wagenmakers2007practical}. We might not include an edge either because the data are too noisy or because the null hypothesis is true; LASSO regularization, like classical significance testing, cannot differentiate between these two reasons. Quantifying evidence for edge weights being zero is still a topic of future research \citep{roadAhead, wetzels2012default}.

\section{Non-normal data}

Common challenges to estimating partial correlation networks relate to the assumption of multivariate normality. The estimation of partial correlation networks is closely related to structural equation modeling \citep{epskampPsychometrika}, and, as such, also requires multivariate normal distributions. Not only does this mean that the marginal distributions must be normal, all relationships between variables must also be linear. But what do we do with non-normal (e.g. ordered categorical) data that are common in psychological data? Several solutions proposed in the structural equation modeling literature may offer solutions to network modeling as well.

\begin{figure}
\centering
 \begin{subfigure}[b]{0.8\linewidth}
	  	\includegraphics[width=1\textwidth,page=1]{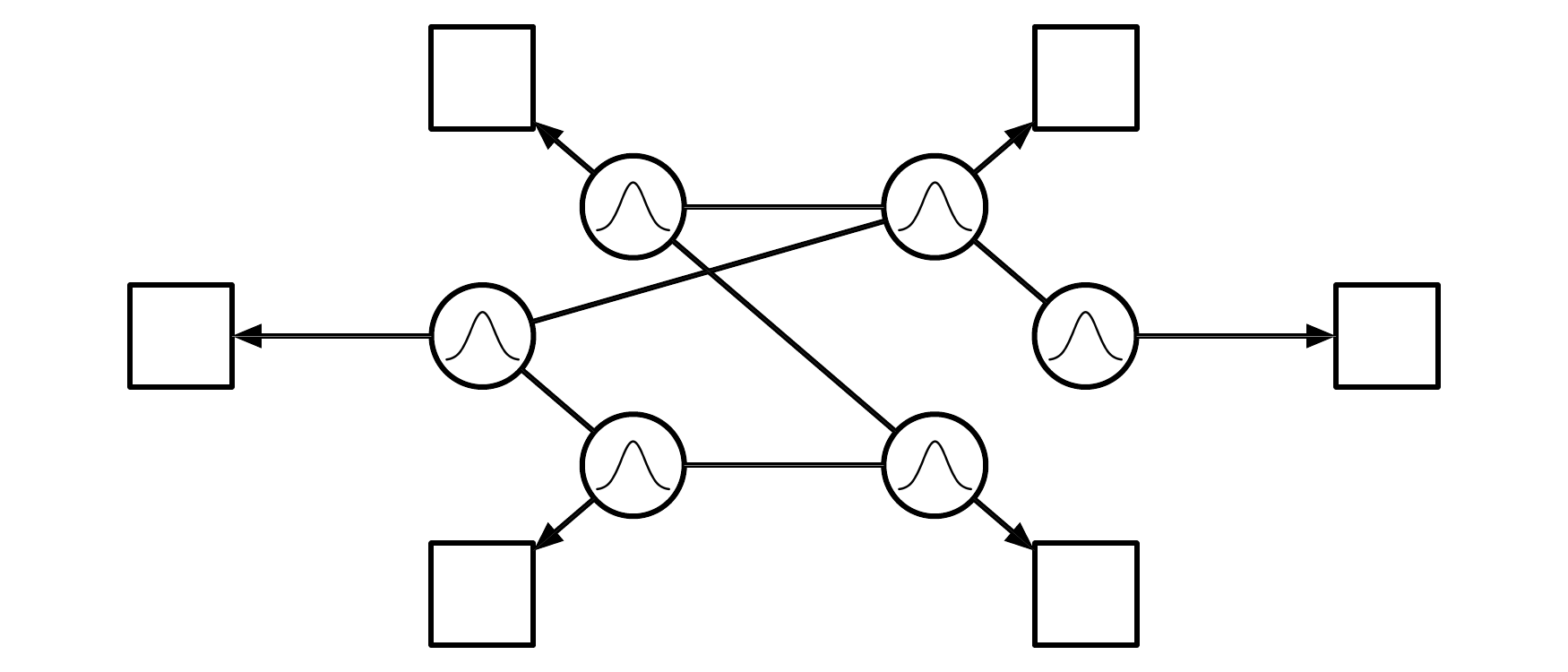}
        \caption{Observed variables (squares) assumed to be transformations of multivariate normal latent variables (circles).}
    \end{subfigure} \\
 \begin{subfigure}[b]{0.8\linewidth}
	  	\includegraphics[width=1\textwidth,page=1]{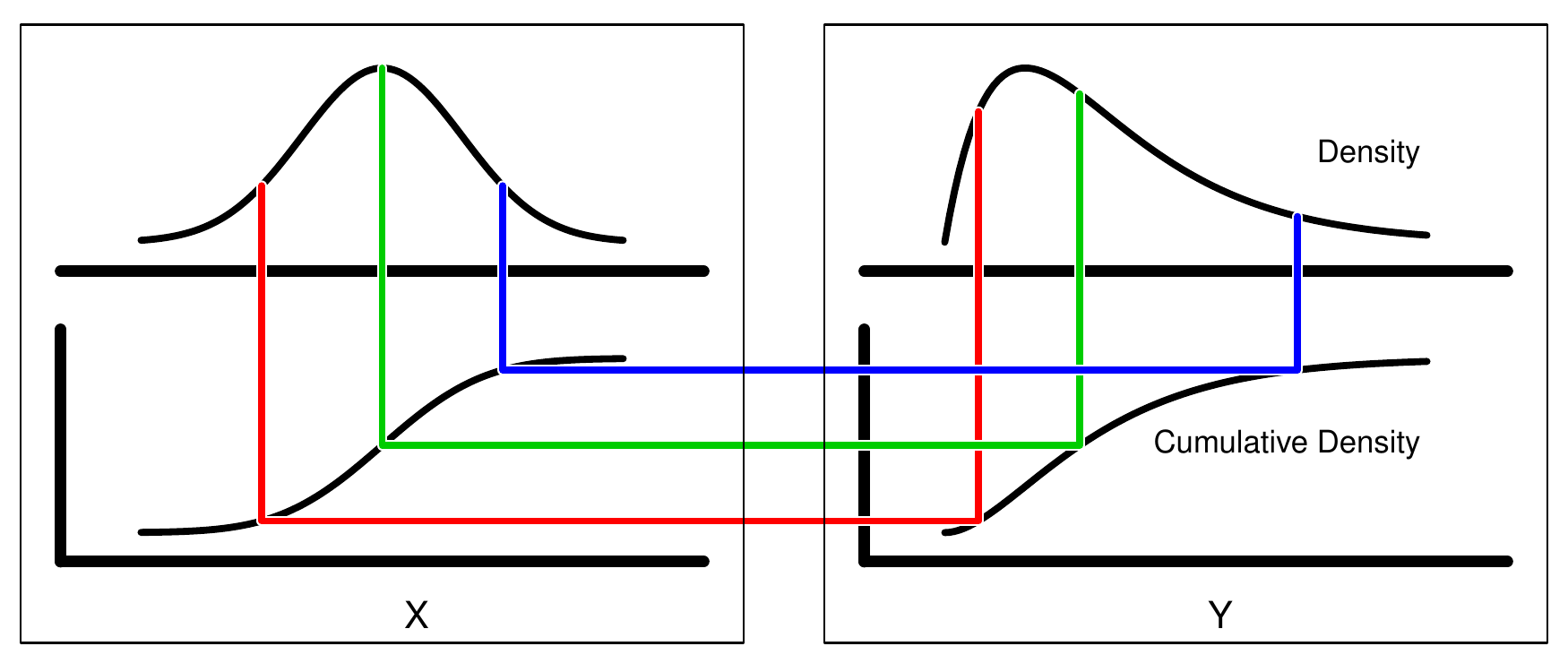}
        \caption{Visualization on how marginal distributions can be used to transform a variable to have a normal marginal distribution.}
    \end{subfigure} \\
     \begin{subfigure}[b]{0.8\linewidth}
	  	\includegraphics[width=1\textwidth,page=1]{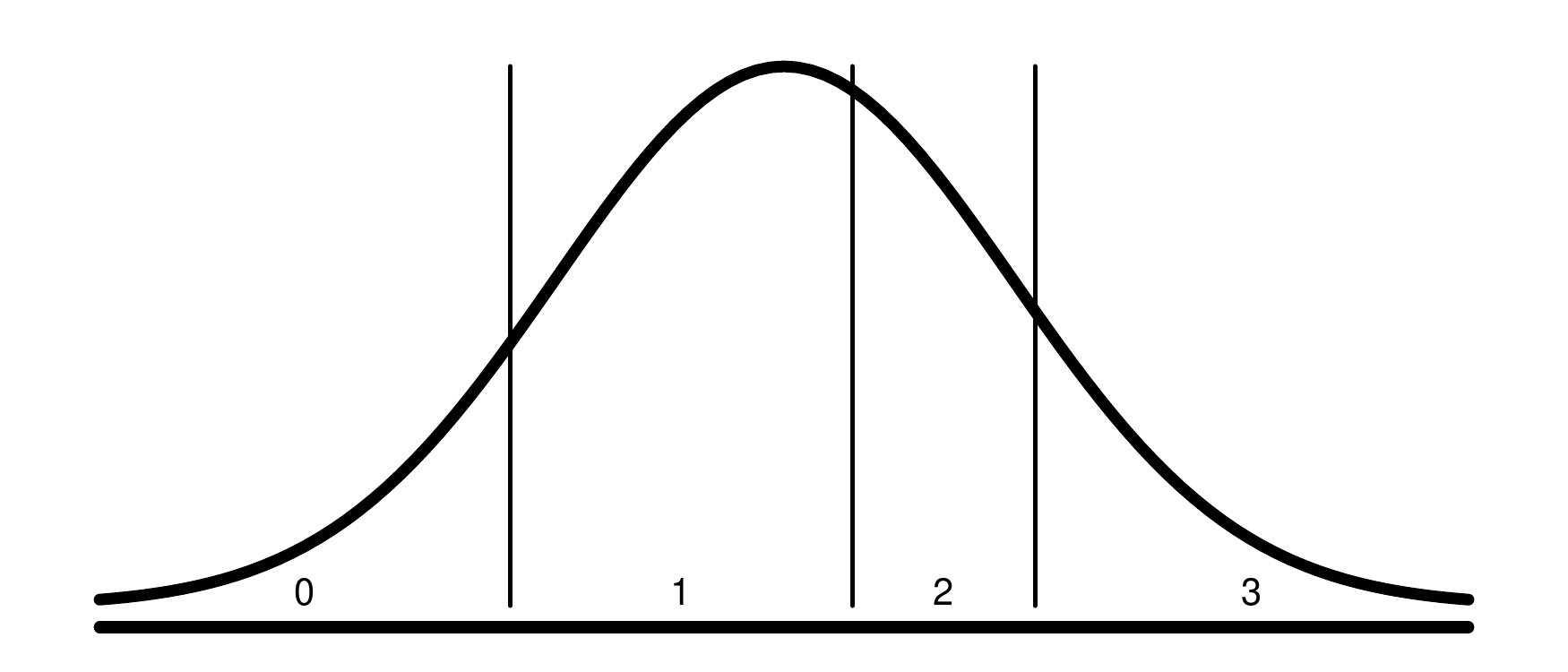}
        \caption{Visualization of a threshold model, used in polychoric and polyserial correlations.}
    \end{subfigure} 
\caption{Methods for relaxing the assumption of multivariate normality.}
\label{primer:fig:nonnormal}
\end{figure}

The assumption of normality can be relaxed by assuming that the observed data are a transformation of a latent multivariate normally distributed system \citep{liu2009nonparanormal}. Figure~\ref{primer:fig:nonnormal}, Panel~(a) shows an example of such a model. In this model, 
squares indicate observed variables, circles indicate normally distributed latent variables and directed arrows indicate monotone (every value is transformed into one unique value, keeping ordinality intact; higher values in the original scale are also higher on the transformed scale) transformation functions. Note that we do not assume measurement error, which could be included by having multiple indicators per latent variable \citep{epskampPsychometrika}. Here, we assume every observed variable indicates one latent variable \citep{muthen1984general}. 

The most common two scenarios are that the observed variables are continuous, or that they consist of ordered categories. When observed variables are continuous, but not normally distributed, the variables can be transformed to have a marginal normal distribution. A powerful method that has been used in network estimation is to apply a \emph{nonparanormal transformation} \citep{liu2009nonparanormal}. This transformation uses the cumulative distributions (encoding the probability that a variable is below some level) to transform the distribution of the observed variable to that of the latent normally distributed variable. Figure~\ref{primer:fig:nonnormal}, Panel~(b) shows a simplified example on how two distributions can be linked by their cumulative distribution. Suppose $X$ is normally distributed, and $Y$ is gamma distributed (potentially skewed). Then, values of $X$ can be mapped to the cumulative distribution by using the probability function (in R: \textVerb{pnorm}). These cumulative probabilities can then be mapped to values of the gamma distribution by using the quantile function (in R: \textVerb{qgamma}). In practice, however, the distribution of $Y$ (top right panel) is not known. The density and cumulative density  of $X$ (left panels), on the other hand, are known, and the cumulative distribution of $Y$ can be estimated by computing the empirical cumulative distribution function (in R: \textVerb{ecdf}). Thus, to map values of $Y$ to values of the normally distributed variable $X$, one needs to estimate a smooth transformation function between the bottom two panels. This is the core of the nonparanormal transformation, which aims to map every unique outcome of a variable (e.g., 1, 2 or 3) to one unique outcome of a standard normal variable (e.g., -1.96, 0, 1.65). The \textVerb{huge.npn} function from the \emph{huge} package \citep{huge} can be used to this end. Important to note is that this transformation assumes smoothly increasing cumulative distributions, and will therefore not work when, only a few possible answering options are present (such as in Likert scales). When the data are binary, the transformed data will still be binary, just using different labels than $0$ and $1$.

When only few item categories are available and the answer options can be assumed to be ordinal \citep{stevens1946theory}, one can make use of threshold functions \citep{muthen1984general} as the data transforming functions. Now, the observed score  is again assumed to be reflective of a latent normally distributed score, but correlations between items can directly be estimated without having to transform the data. An example of such a threshold function is shown in Figure~\ref{primer:fig:nonnormal}, Panel~(c). In this panel, three thresholds are estimated to accommodate four answering categories ($0$, $1$, $2$ or $3$). The normal distribution corresponds to the latent item score and vertical bars correspond to the thresholds; a person with a latent score below the first would score a $0$, a person with a latent score between the first and second threshold would score a $1$, etc. After the thresholds are estimated, the correlations between latent variables can be estimated pairwise. These are termed \emph{polychoric} correlations when both variables are ordinal \citep{olsson1979maximum}, or \emph{polyserial} correlations when only one of the two variables is ordinal \citep{olsson1982polyserial}.  The \textVerb{lavCor} function from the \emph{lavaan} package \citep{lavaan} can be used to compute polychoric and polyserial correlations, which can subsequently be used as input to the glasso algorithm \citep{qgraphsims}. Regularized partial correlations using glasso with EBIC model selection based on polychoric correlations has become standard when estimating psychopathology networks due to the high prevalence of ordered-categorical data. An important limitation is that these methods rely on an assumption that the latent variables underlying the observed ordinal variables are normally distributed, which might not be plausible. For example, some psychopathological symptoms, such as suicidal ideation, might plausibly have a real ``zero'' point — the absence of a symptom. Properly handling such variables is still a topic of future research \citep{roadAhead}. 
 
When data are binary, one could also use tetrachoric and biserial correlations (special cases of polychoric and polyserial correlations respectively). However, these data would not be best handled using partial correlation networks because of the underlying assumption of normality. When all variables are binary, one can estimate the Ising Model using the \emph{IsingFit} R package \citep{IsingFit}. The resulting network has a similar interpretation as partial correlation networks, and is also estimated using LASSO with EBIC model selection \citep{van2014new}. When the data consist of both categorical and continuous variables, the state-of-the-art network model is termed the Mixed Graphical Model, which is implemented in the \emph{mgm} package \citep{mgm}, also making use of LASSO estimation with EBIC model selection.

\section{Example}

In this section, we estimate a network based on the data of 221 people with a sub-threshold post-traumatic stress disorder (PTSD) diagnosis. The network features 20 PTSD symptoms. A detailed description of the dataset can be found elsewhere \citep{armour2017network}, and the full R code for this analysis can be found in the supplementary materials.\footnote{We performed these analyses using \emph{R} version 3.5.0, \emph{bootnet} version 1.0.1 and \emph{qgraph} version 1.4.4, using OSX version 10.11.6.}

\begin{figure*}
\centering
 \begin{subfigure}[b]{0.33\linewidth}
	  	\includegraphics[width=1\textwidth,page=1]{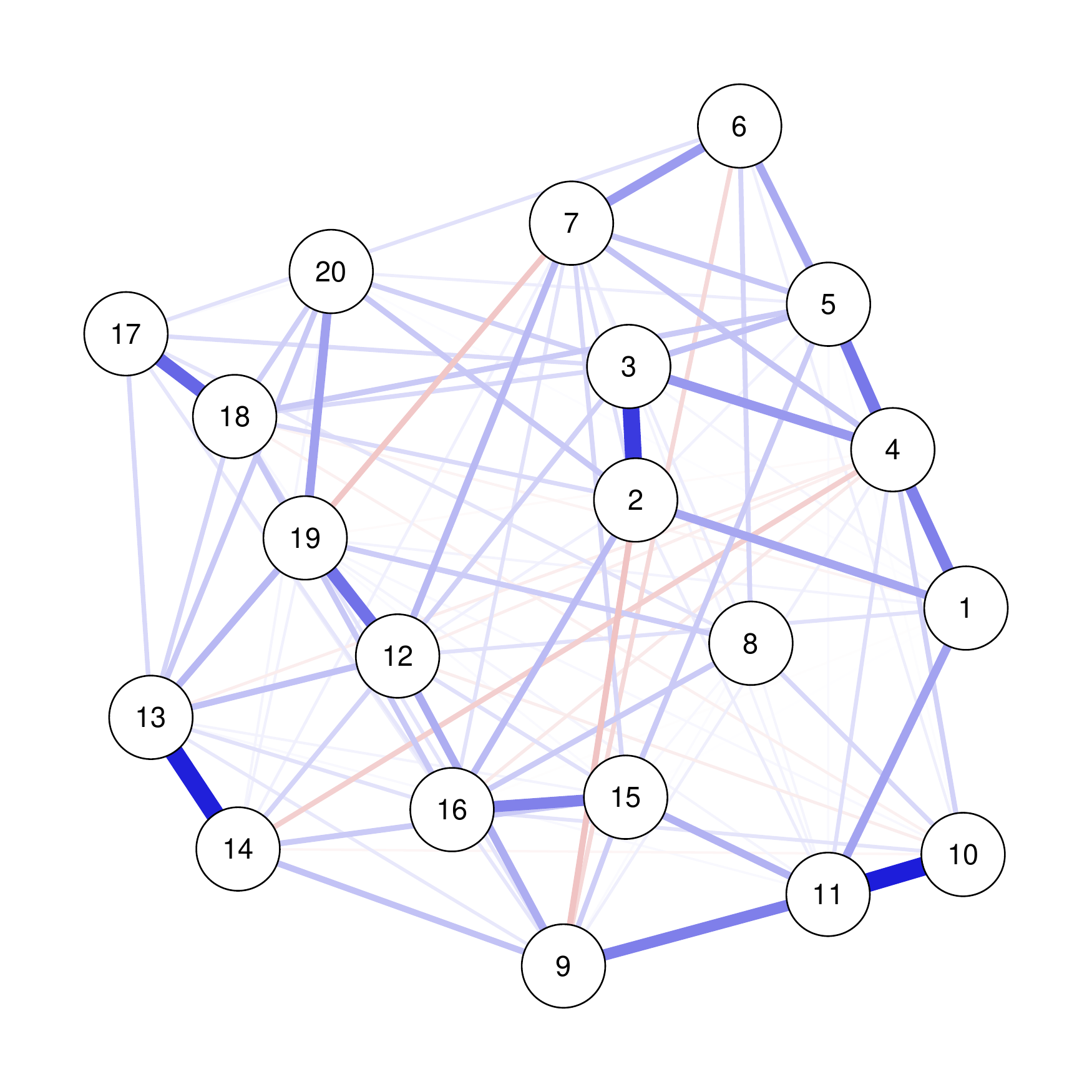}
        \caption{$\gamma = 0$\label{primer:fig:3a}}
    \end{subfigure}  
\begin{subfigure}[b]{0.33\linewidth}
	  	\includegraphics[width=1\textwidth,page=2]{Fig3.pdf}
        \caption{$\gamma = 0.25$\label{primer:fig:3b}}
    \end{subfigure} 
 \begin{subfigure}[b]{0.33\linewidth}
	  	\includegraphics[width=1\textwidth,page=3]{Fig3.pdf}
        \caption{$\gamma = 0.5$\label{primer:fig:3c}}
    \end{subfigure}
\caption{Partial correlation networks estimated on responses of 221 subjects on 20 PTSD symptoms, with increasing levels of the LASSO hyperparameter $\gamma$ (from left to right: Panel~(a) = 0, Panel~(b) = 0.25, Panel~(c) = 0.5).}
\label{primer:fig:3}
\end{figure*}

The following R code performs regularized estimation of a partial correlation network using EBIC selection \citep{foygel2010extended}. This methodology has been implemented in the \textVerb{EBICglasso} function from the \emph{qgraph} package \citep{jssv048i04}, which in turn utilizes the \emph{glasso} package for the glasso algorithm \citep{glasso}. A convenient wrapper around this (and several other network estimation methodologies such as the Ising Model and the Mixed Graphical Model) is implemented in the \emph{bootnet} package (see \citealt{bootnetpaper} for an extensive description), which we use here in order to perform (a) model estimation, (b) a priori sample size analysis, and (c) post-hoc accuracy and stability analysis. This code assumes the data is present in R under the object name \textVerb{data}.
\begin{verbatim}
library("bootnet")
results <- estimateNetwork(
  data,
  default = "EBICglasso",
  corMethod = "cor_auto",
  tuning = 0.5)
\end{verbatim}
In this code, \textVerb{library("bootnet")} loads the package into R, and the \textVerb{ default = "EBICglasso"} specifies that the \textVerb{EBICglasso} function from \emph{qgraph} is used. The \textVerb{corMethod = "cor\_auto"} argument specifies that the \textVerb{cor\_auto} function from \emph{qgraph} is used to obtain the necessary correlations. This function automatically detects ordinal variables (variables with up to 7 unique integer values) and uses the \emph{lavaan} package \citep{lavaan} to estimate polychoric, polyserial and Pearson correlations. Finally, the \textVerb{tuning = 0.5} argument sets the EBIC hyperparameter, $\gamma$, to $0.5$. After estimation, the network structure can be obtained using the code:
\begin{verbatim}
results$graph
\end{verbatim}
and the network can be plotted using the plot method of \emph{bootnet} using the code: 
% EF you don't need to cite qgraph everytime you mention it, sae with bootnet.
\begin{verbatim}
plot(results)
\end{verbatim}
This function uses the  \textVerb{qgraph} function from the \emph{qgraph} package to draw the network \citep{jssv048i04}.\footnote{Any argument used in this plot method is used in the underlying call to \textVerb{qgraph}. The \emph{bootnet} plot method has three different default arguments than \emph{qgraph}: (1) the \textVerb{cut} argument is set to zero, the \textVerb{layout} argument is set to \textVerb{"spring"}, and the \textVerb{theme} argument is set to \textVerb{"colorblind"}. For more details on the these arguments and other ways in which \emph{qgraph} visualizes networks we refer the reader to \citet{jssv048i04} and the online documentation at https://CRAN.R-project.org/package=qgraph.}
% By default, edge weights are visualized using green or red edges that are drawn more saturated and wider the stronger the absolute value of the corresponding partial correlation coefficient. With more than $20$ nodes in a network, \emph{qgraph} automatically splits scaling in width and saturation of edges based on some cutoff value: edges with absolute edge weights (partial correlations) under the cutoff value are drawn as thin edges that vary in saturation, and edges with absolute edge weights above the cutoff value are drawn as fully saturated edges that vary in width. We recommend to not use this feature for visualizing regularized partial correlation networks, and instead disable the cutoff by using the argument \textVerb{cut = 0}. The argument \textVerb{layout} specifies the node placement. Setting \textVerb{layout = "spring"} 
By default, edges are drawn using a colorblind-friendly theme (blue edges indicate positive partial correlations and red edges indicate negative partial correlations). Nodes are placed using an modified version of the Fruchterman-Reingold algorithm \citep{fruchterman1991graph} for weighted networks \citep{jssv048i04}. This algorithm aims to place nodes in an informative way by positioning connected nodes close to each other. A downside of the Fruchterman-Reingold algorithm is that it can behave chaotically: every input will lead to one exact output, but small differences in the input (e.g., a difference of $0.01$ in an edge weight or using a different computer architecture) can lead to an entirely different placement of nodes (nodes will likely be placed about the same distance from one-another, but might be placed on a different side of the plotting area). Thus, the eventual placement cannot be interpreted in any substantial way, and might differ substantially between two networks even when there are only very small differences in the network structures. To compare two networks, one should constrain the layout to be equal for both networks. One way to do so is by using \textVerb{averageLayout} from the \emph{qgraph} package, which was used in drawing Figure~\ref{primer:fig:3}\footnote{See online supplementary materials for exact R codes.}.

% \textVerb{estimateNetwork} function is used to estimate the network structure. The argument  \textVerb{corMethod = "cor_auto"}  the \textVerb{cor\_auto} function detects ordinal variables (variables with up to 7 unique integer values) and uses the \emph{lavaan} package \citep{lavaan} to estimate polychoric, polyserial and Pearson correlations. The \textVerb{qgraph} function estimates and plots the network structure. The argument \textVerb{graph} specified that we want to use the glasso algorithm with EBIC model selection, the argument \textVerb{sampleSize} specifies the sample size of the data, the argument \textVerb{layout} specifies the node placement and the argument tuning specified the EBIC hyperparameter. The hyperparameter is here set to $0.5$, which is also the current default value used in \emph{qgraph}. For more control on the estimation procedure, one can use the \textVerb{EBICglasso} function, which is automatically called when using \textVerb{qgraph(..., graph = "glasso")}. Finally, the estimated weights matrix can be obtained either directly using \textVerb{EBICglasso} or by using the \textVerb{getWmat} function on the output of \textVerb{qgraph}:
%\begin{verbatim}
%getWmat(graph)
%\end{verbatim}

Figure~\ref{primer:fig:3} shows the resulting network estimated under three different values of $\gamma$: $0$, $0.25$, and $0.5$. Table~\ref{primer:nodeLabels} shows the description of the nodes. As expected, the network with the largest hyperparameter has the fewest edges: the networks feature 105 edges with $\gamma=0$, 95 edges with $\gamma=0.25$, and 87 edges with $\gamma=0.5$.

\begin{table}[]
\centering
\caption{Description of nodes shown in Figure~\ref{primer:fig:3}}
\label{primer:nodeLabels}
\begin{tabular}{lllll}
Node & Description                   \\
\hline
1    & Intrusive Thoughts           \\
2    & Nightmares                   \\
3    & Flashbacks                   \\
4    & Emotional cue reactivity     \\
5    & Psychological cue reactivity \\
6    & Avoidance of thoughts        \\
7    & Avoidance of reminders       \\
8    & Trauma-related amnesia       \\
9 & Negative beliefs \\
10 & Blame of self or others \\
11 & Negative trauma-related emotions \\
12 & Loss of interest \\
13 & Detachment \\
14 & Restricted affect \\
15 & Irritability/anger \\
16 & Self-destructive/reckless behavior \\
17 & Hypervigilance \\
18 & Exaggerated startle response \\
19 & Difficulty concentrating \\
20 & Sleep disturbance \\
\hline
\end{tabular}
\end{table}

\begin{figure}
\centering
\includegraphics[width=0.9\linewidth]{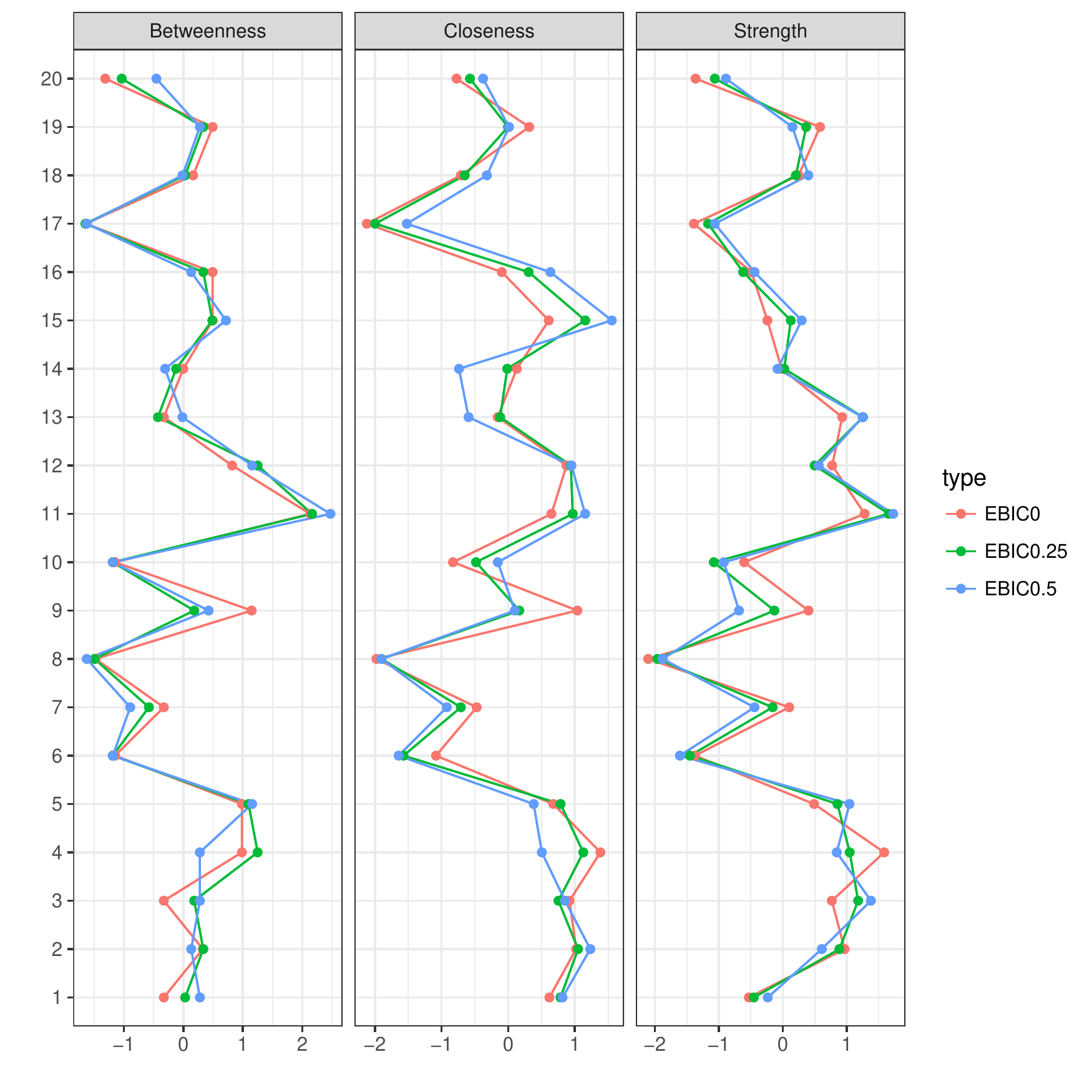}
\caption{ Closeness, betweenness, and degree centrality of the three networks described in Figure~\ref{primer:fig:3} with increasing levels of the LASSO hyperparameter $\gamma$. Centrality indices are plotted using standardized $z$-scores in order to facilitate interpretation.}
\label{primer:fig:4}
\end{figure}

We can further investigate how important nodes are in the network using measures called centrality indices. These indices can be obtained as followed:
\begin{verbatim}
	centrality(results)
\end{verbatim}
This code provides three commonly used centrality indices: \emph{node strength}, which takes the sum of absolute edge weights connected to each node, \emph{closeness}, which takes the inverse of the sum of distances from one node to all other nodes in the network, and \emph{betweenness}, which quantifies how often one node is in the shortest paths between other nodes. A more extensive overview of these measures and their interpretation is described elsewhere \citep{costantini2015state, bootnetpaper, opsahl2010node}. All measures indicate how important nodes are in a network, with higher values indicating that nodes are more important. Figure~\ref{primer:fig:4} is the result of the function \textVerb{centralityPlot} and shows the centrality of all three networks shown in Figure~\ref{primer:fig:3}. For a substantive interpretation of the network model obtained from this dataset we refer the reader to \citet{armour2017network}.

\section{Sample size selection and replicability}

An increasingly important topic in psychological research is the replicability of results \citep{opensciencecollaboration2015}. High-dimensional exploratory network estimation, as presented in this tutorial paper, lends itself to generating many different measures (e.g., edge weights, network structures, centrality indices) that may or may not replicate or generalize across samples. Recent work has put the importance of replicability in network modeling of psychological data in the spotlight (\citealt{bootnetpaper, friedreview, friedchallenge,fried2017replicability}; \citealt{forbes}, but see also \citealt{forbescommentary}). However, in is not easy to determine the replicability of an estimated network. Many factors can influence the stability and accuracy of results, such as the sample size, the true network structure and other characteristics of the data\footnote{For example, \citet{forbescommentary} show how data-imputation strategies can lead to unstable edge parameters even at large sample size.}. Even when a network is estimated stably, measures derived from the network structure (e.g., graph theoretical measures such as centrality metrics) might still not be interpretable. For example, all nodes in the true network shown in Figure~\ref{primer:fig:1} have exactly the same betweenness (0, all shortest paths do not go via third nodes). Thus, any differences in betweenness in estimated networks are due to chance, regardless of sample size.

We therefore recommend sample size analyses both before and after collecting the data for analysis. A priori sample size analyses let researchers know if the sample size is appropriate for the expected network structure, and post-hoc stability analyses provide researchers with information about the stability of their results. We describe a priori sample size analysis in detail in the next section, which has not been done before in the psychological network literature, and then summarize post-hoc stability analyses that are explicated in detail elsewhere \citep{bootnetpaper}.

\subsection{A priori sample size analysis}

An important consideration for any statistical analysis is the sample size required for an analysis, which is often referred to as \emph{power analysis} \citep{cohen1977statistical}. To perform such an analysis, one needs to have a prior expectation of the effect size---the expected strength of the true effect. In network modeling, the analogy to an expected effect size is the expected weighted network: a high-dimensional interplay of the network structure (zero and non-zero edges) and the strength of edges (the weight of the non-zero edges). For a partial correlation network of $P$ nodes, one needs to have a prior expectation on $P (P-1) / 2$ parameters (edges) to estimate how well edges or any descriptive statistics derived from the network structure, such as centrality indices, can be estimated stably given a certain sample size.\footnote{Other network models, such as the Ising model, also require a prior expectation for the $P$ intercepts. The partial correlation network does not require intercepts as data can be assumed centered.}

When estimating a network structure, three properties are of primary interest \citep{van2014new}:
\begin{itemize}
\item \emph{Sensitivity}: Also termed the true-positive rate, the proportion of edges present in the true network that were detected in the estimated network.
\item \emph{Specificity}: Also termed the true-negative rate, the proportion of missing edges in the true network that were also detected correctly to be absent edges in the estimated network.
\item The \emph{correlation} between edge weights of the true network and edge weights of the estimated network, or between centrality estimates based on the true network and centrality estimates based on the estimated network.
\end{itemize}
A researcher wants sensitivity to increase with sample size and preferably to be high (although a moderate sensitivity can be acceptable as that at least indicates the strongest edges are discovered). When specificity is low, the estimation procedure mistakingly detects many edges that are not present in the true network (false positives). As a result, we argue that researchers always want high specificity. Finally, the correlation indicates how well the true network structure and the estimated network structure mimic one-another. Especially when a researcher is interested in analyzing the network structure as a whole (e.g., for shortest paths analyses), the researcher wants this to be high. In addition to this correlation, the correlation between between centrality indices of the true network and the estimated network might also be of interest, which can be low even though the edge weights are estimated accurately (e.g., when centrality does not differ in the true network, such as betweenness in Figure~\ref{primer:fig:1}). 

Simulation studies have shown that LASSO regularized network estimation generally results in a high specificity, while sensitivity and correlation increases with sample size \citep{qgraphsims,van2014new,foygel2010extended}. This means that whenever LASSO regularization is used, one can interpret edges that are discovered by the method as likely to represent edges in the true network, but should take into account that the method might not discover some true edges. Unfortunately, the precise values of sensitivity, specificity and different correlations are strongly influenced by the expected network structure, similar to how the expected effect size influences a power analysis. As a result, judging the required sample size is far from trivial, but has been called for multiple times in the recent literature \citep{bootnetpaper,friedchallenge}. 

We recommend three ways forward on this issue: (1) more research estimating network models from psychological data will make clear what one could expect as a true network structure, especially if researchers make the statistical parameters of their network models publicly available; (2) researchers should simulate network models under a wide variety of potential true network structures, using different estimation methods; (3) researchers should simulate data under an expected network structure to gain some insight in the required sample size. To aid researchers in (2) and (3), we have implemented the \textVerb{netSimulator} function in the \emph{bootnet} package, which can be used to flexibly set up simulation studies assessing sample size and estimation methods given an expected network structure.

\begin{figure*}
\centering	  	
	\includegraphics[width=0.49\linewidth,page=1]{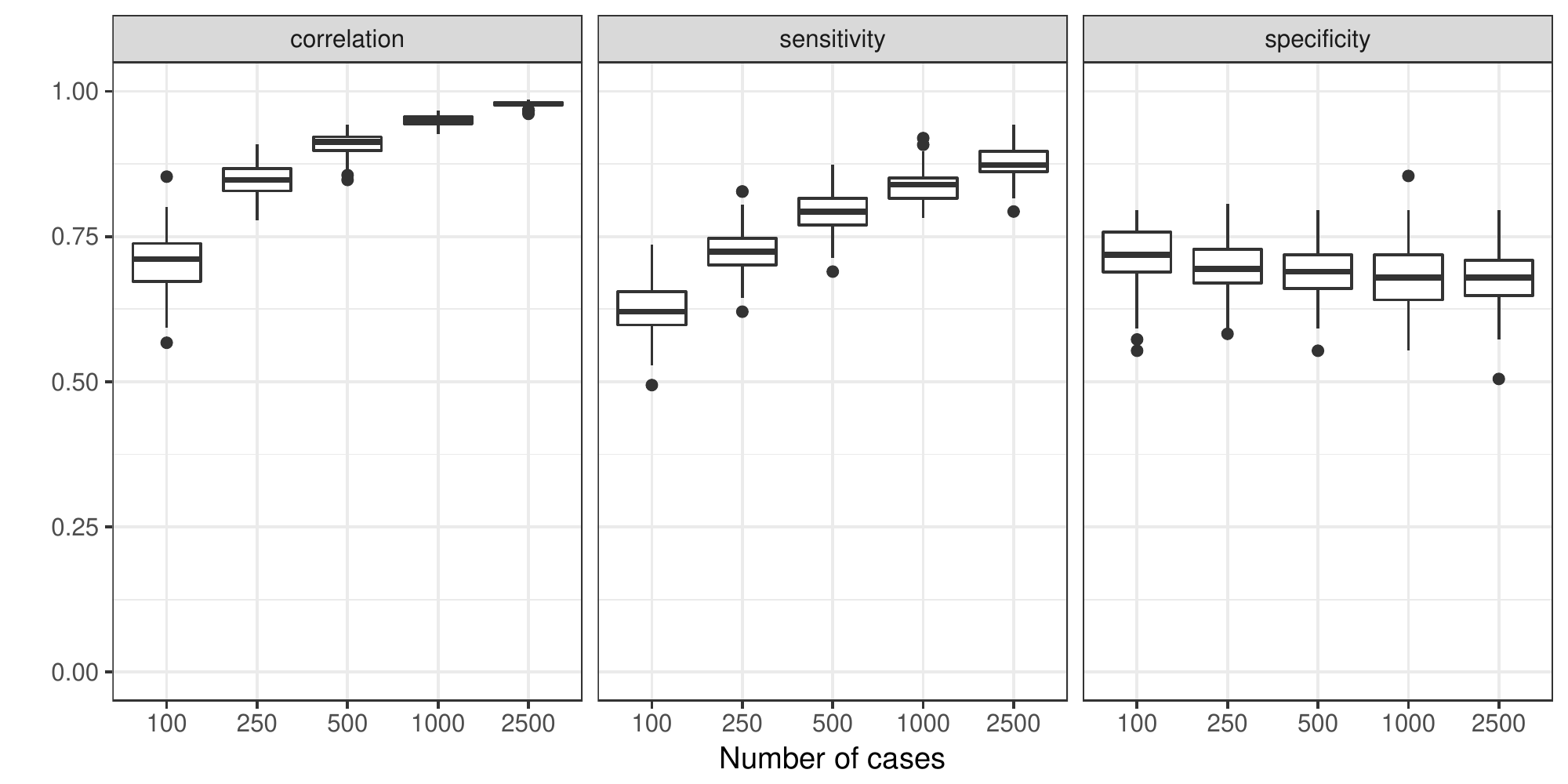} 
 	\includegraphics[width=0.49\linewidth,page=2]{Fig6.pdf}
\caption{Simulation results using the estimated refitted PTSD network as true network structure. The top panel shows the sensitivity (true positive rate), specificity (true negative rate) and correlation between true and estimated networks, and the bottom panel shows the correlation between true and estimated centrality indices.}
\label{primer:fig:6}
\end{figure*}

The \textVerb{netSimulator} function can simulate data under a given network model and expected network structure. Since partial correlation networks feature many parameters, and the field of estimating these models is still young, researchers cannot be expected to have strong theoretical expectations on the network structure. One option is to simulate data under the parameters of a previously published network model, which can be obtained by re-analyzing the data of the original authors or, if the data are not available, asking the original authors to send the adjacency matrix encoding the edge weights. Below, we will conduct such a simulation study by using the estimated network structure in Figure~\ref{primer:fig:3}, Panel~(c) as the simulation baseline. 
Simulating data under LASSO regularized parameters, however, poses a problem in that these parameters will be biased towards zero due to shrinkage, and therefore might imply a weaker effect than can be expected. To accommodate this, we can first fit the model by using LASSO to obtain a network structure (i.e. which edges are present), and then \emph{refit} a model with only those edges without LASSO regularization (see also \citealt{epskampPsychometrika} on confirmatory partial correlation network analysis). This can be done by using the \textVerb{refit} argument in \textVerb{estimateNetwork}:
\begin{verbatim}
network <- estimateNetwork(
  data,
  default = "EBICglasso",
  corMethod = "cor_auto",
  tuning = 0.5,
  refit = TRUE)
\end{verbatim}
Next, a simulation study can be performed using the following R code:
\begin{verbatim}
simRes <- netSimulator(network$graph, 
             dataGenerator = ggmGenerator(
             	ordinal = TRUE, nLevels = 5),
             default = "EBICglasso",
             nCases = c(100,250,500,1000,2500),
             tuning = 0.5,
             nReps = 100,
             nCores = 8
             )
\end{verbatim} 
The \textVerb{netSimulator} can use any argument of \textVerb{estimateNetwork}, with a vector of options describing multiple conditions are estimated (e.g., \textVerb{tuning = c(0.25, 0.5)} would vary the tuning parameter). The first argument is a weights matrix encoding an expected network (or a list with a weights matrix and intercepts vector for the Ising Model which is not needed for partial correlation networks), the \textVerb{dataGenerator} argument specifies the data generating process (can be ignored for non-ordinal data), \textVerb{nCases} encodes the sample size conditions, \textVerb{nReps} the number of repetitions per condition, and \textVerb{nCores} the number of computer cores to use. Next, results can be printed:
\begin{verbatim}
simRes
\end{verbatim}
or plotted:
\begin{verbatim}
plot(simRes)
plot(simRes,
 yvar = c("strength","closeness","betweenness"))
\end{verbatim}
Figure~\ref{primer:fig:6} shows the corresponding plots. These plots may be used to gain a rough insight into the required sample size, based on the requirements of the researcher. For example, $N = 250$ achieves a correlation between the 'true' and estimated networks above $0.8$ for edge weights and strength, and above $0.7$ for sensitivity. Noteworthy is that specificity is moderate, but not as high as in other studies \citep{qgraphsims,van2014new,foygel2010extended}, possibly a result of the true network structure used being very sparse ($54\%$ of the edges were zero in the generating network).

\subsection{Post-hoc stability analysis}

After estimating a network, bootstrapping methods \citep{chernick2011bootstrap,efron1979bootstrap} can be used to gain insight into the accuracy and stability of the network parameters and descriptive statistics based on the estimated network structure (e.g., centrality indices). These are extensively discussed by \citet{bootnetpaper}, including a tutorial on how to perform these analyses using the \emph{bootnet} package. In short, \emph{bootnet} can be used to perform several types of bootstraps using the original data and the estimation method. The two most important methods are:
\begin{verbatim}
boot1 <- bootnet(results, nCores = 8, 
	nBoots = 1000, type = "nonparametric")
boot2 <- bootnet(results, nCores = 8, 
	nBoots = 1000, type = "case")
\end{verbatim}
The first bootstrap is a non-parametric bootstrap (using resampled data with replacement), which can be used to construct confidence intervals around the regularized edge weights \citep{hastie2015statistical} and perform significance tests on the difference between different edge weights (e.g. comparing edge $A$ -- $B$ with edge $A$ -- $C$) and different centrality indices (e.g., comparing node strength centrality of node A vs.\ node B). Confidence intervals can \emph{not} be constructed for centrality indices (see the supplementary materials of \citealt{bootnetpaper}). To assess the stability of centrality indices, one can perform a case-dropping bootstrap (subsampling without replacement). Based on these bootstraps, the steps from \citet{bootnetpaper} can be followed to create several plots, which we include for the network in Figure~\ref{primer:fig:3}, Panel~(c)
in the supplementary files to this paper. The plots show sizable sampling variation around the edge weights and a poor stability for closeness and betweenness. Strength was more stable, although not many nodes differed from each other significantly in strength. The results of the case-dropping bootstrap can also be summarized in a coefficient, the $CS$-coefficient (correlation stability), which quantifies the proportion of data that can be dropped to retain with $95\%$ certainty a correlation of at least $0.7$ with the original centrality coefficients. Ideally this coefficient should be above $0.5$, and should be at least above $0.25$. Strength was shown to be stable ($CS(\mathrm{cor} = 0.7) \approx 0.516$) while closeness ($CS(\mathrm{cor} = 0.7) \approx 0.204$) and betweenness ($CS(\mathrm{cor} = 0.7) \approx 0.05$) were not. Thus, the post-hoc analysis shows that the estimated network structure and derived centrality indices should be interpreted with some care for our example network of PTSD symptoms.

\section{Common Problems and Questions}

\begin{figure*}
\centering
 \begin{subfigure}[b]{0.45\linewidth}
	  	\includegraphics[width=1\textwidth,page=1]{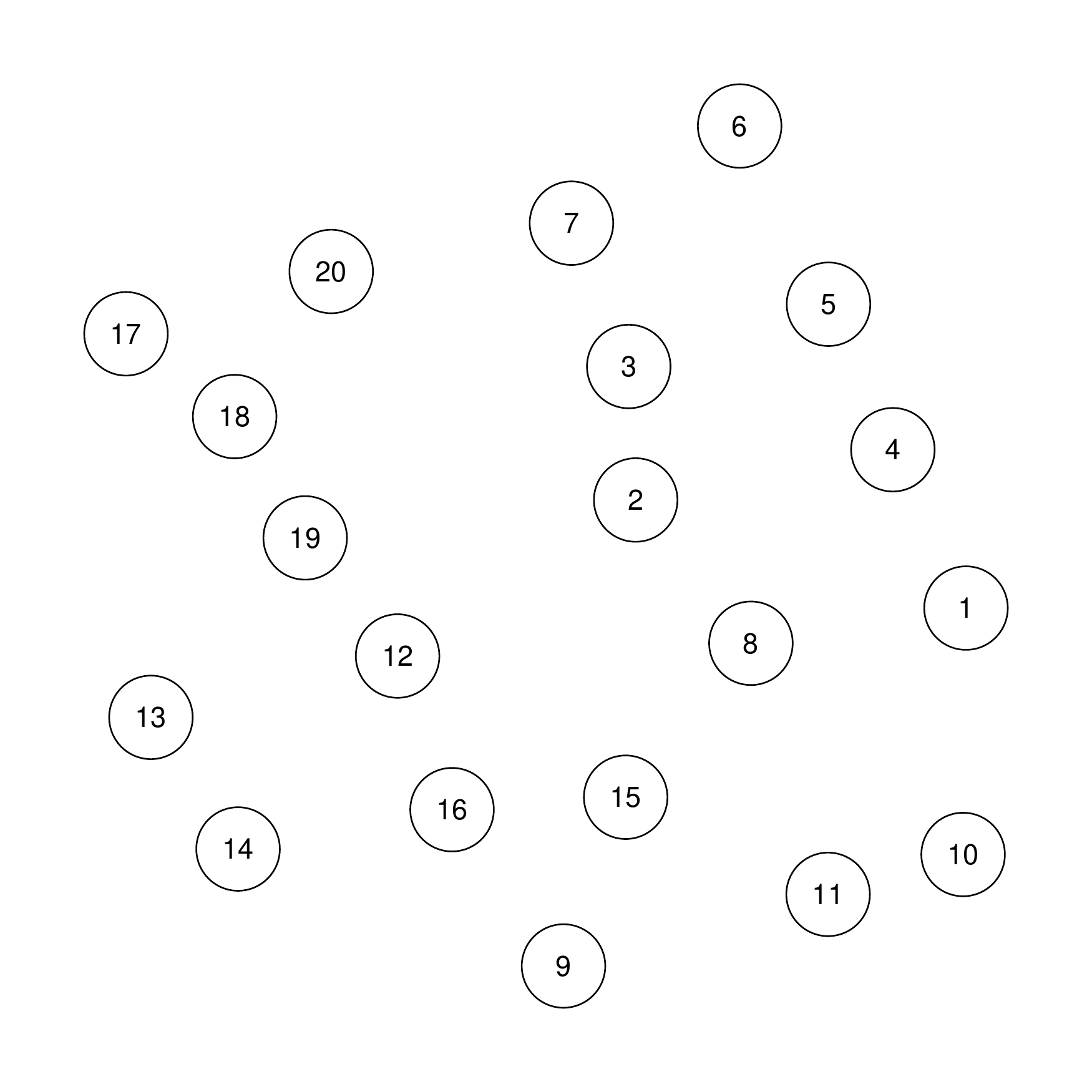}
        \caption{$\gamma = 0.5$\label{primer:fig:5a}}
    \end{subfigure}
 \begin{subfigure}[b]{0.45\linewidth}
	  	\includegraphics[width=1\textwidth,page=2]{Fig5.pdf}
        \caption{$\gamma = 0$\label{primer:fig:5b}}
    \end{subfigure}
\caption{Network of 20 PTSD symptoms. Instead of the full data like in Figure~\ref{primer:fig:3} (221 subjects), only 50 subjects were used. Panel~(a): LASSO hyperparameter $\gamma$ set to the default of 0.5; Panel~(b): $\gamma$ set to 0 for discovery.}
\label{primer:fig:5}
\end{figure*}

\paragraph{Difficulties in interpreting networks} Regularized networks can sometimes lead to network structures that are hard to interpret. Here, we list several common problems and questions encountered when estimating and interpreting these models, and try to provide potential ways forward. 
\begin{enumerate}
\item The estimated network has no or very few edges. This can occur in the unlikely case when variables of interest do not exhibit (partial) correlations. More likely, it occurs when the sample size is too low for the number of nodes in the network. The EBIC penalizes edge weights based on sample size to avoid false positive associations, which means that with increasing sample size, the partial correlation network will be more and more similar to the regularized partial correlation network. With smaller $N$ fewer edges will be retained. Figure~\ref{primer:fig:5}, Panel~(a) shows a network estimated on the same data as Figure~\ref{primer:fig:3}, but this time with only 50 instead of the 221 participants: it is devoid of any edges. A way to remediate this problem is by setting the hyperparameter lower (e.g., 0; see Figure~\ref{primer:fig:5}, Panel~(b)), but note that this increases the likelihood that the network will contain spurious edges. An alternative solution is to reduce the variables of interest and estimate a network based on a subset of variables, because fewer nodes mean that fewer parameters are estimated. However, doing so would lead one to not use all the available data, and might lead to failing to condition on relevant nodes.
\item The network is densely connected (i.e., many edges) including many unexpected negative edges and many implausibly high partial correlations (e.g., higher than 0.8). As the LASSO aims to remove edges and returns a relatively sparse network, we would not expect densely connected networks in any data that are not extremely large. In addition, we would not expect many partial correlations to be so high, as (partial) correlations above 0.8 indicate near-perfect collinearity between variables. These structures can occur when the correlation matrix used as input is not \emph{positive definite}, which can occur when a sample is too small, or when estimating polychoric correlations. Just as a variance has to be positive, a variance--covariance matrix has to be positive-definite (all eigenvalues higher than zero) or at least positive semi-definite (all eigenvalues at least zero). When a covariance matrix is estimated pairwise, however, the resulting matrix is not guaranteed to be positive-definite or positive-semi-definite. Polychoric correlation matrices are estimated in such a pairwise manner. In case of a non-positive definite correlation matrix, \textVerb{cor\_auto} will warn the user when it estimates a non-positive definite correlation matrix and attempt to correct for this by searching for a nearest positive definite matrix. This matrix, however, can still lead to very unstable results. When the network looks very strongly connected with few (if any) missing edges and partial correlations near $1$ and $-1$, the network structure is likely resulting from such a problem and should not be interpreted. We suggest that researchers always compare networks based on polychoric correlations with networks based on Spearman correlations (they should look somewhat similar) to determine if the estimation of polychoric correlations is the source of this problem. % EF maybe go over the above section one time briefly, could be streamlined a tiny bit. 
\item While in general the graph looks as expected (i.e., relatively sparse), some edges are extremely high and/or unexpectedly extremely negative. This problem is related to the previous problem. The estimation of polychoric correlations relies on the pairwise crossing of variables in the dataset. When the sample size is relatively low, some cells in the item by item frequency table can be low or even zero (e.g., nobody was observed that scored a 2 on one item and a 1 on another item). The estimation of polychoric correlations is based on these frequency tables and is biased whenever an expected frequency is too small (i.e., below 10; \citealt{olsson1979maximum}). Low frequencies  can thus lead to biased polychoric correlations, which can compound to large biases in the estimated partial correlations. Another situation in which one might obtain low frequencies is when the scores are highly skewed \citep{rigdon1991performance}, which unfortunately often is the case in psychopathology data. Again, the network based on polychoric correlations should be compared to a network based on Spearman correlations. Obtaining very different networks indicates that the estimation of the polychoric correlations may not be trustworthy.
\item 
A network has negative edges where the researcher would expect positive ones. This can occur when one conditions on a \emph{common effect} \citep{pearl2000causality}. Suppose one measures three variables: psychology students’ grades on a recent statistics test, their motivation to pass the test, and the easiness of the test \citep{koller2009probabilistic}. The grade is likely positively influenced by both test easiness and student motivation, and we do not expect any correlation between motivation and easiness: knowing a student is motivated does not help us predict how difficult a professor makes a test. However, we can artificially induce a negative partial correlation between motivation and easiness by conditioning on a common effect: if we know an unmotivated student obtained an A, we now \emph{can} expect that the test must have been very easy. These negative relationships can occur when common effect relationships are present, and unexpected negative relationships might indicate common effect structures. 

Another way these unexpected negative relationships can occur is if the network is based on a subsample, defined by a function on the observed variables. This is because taking a subsample based on a function of the observed variables is the same as conditioning on a common effect \citep{muthen1989factor}. For example, a function of the observed variables might be the sum-score. When using this sum-score to select people to include in the analysis (e.g., to investigate the network structure of subjects with severe symptoms compared to subjects with less severe symptoms), then that subsample is derived by conditioning on the sumscore (e.g., only people with a sumscore above 10 are included). This will lead to spurious negative edges in the expected network structure \citep{muthen1989factor}. Results based on such subsamples should be interpreted with care.  In general, this poses a somewhat curious problem: on the one hand, we want to include as many variables as possible; on the other hand, we want to avoid controlling for (i.e. condition on) common effects. Important to note is that one would \emph{not} expect negative partial correlations to occur if the common cause model is true and all variables are scored such that factor loadings are positive \citep{holland1986conditional}, as such negative relationships where one would expect positive ones can be of particular interest to the researcher.  
\end{enumerate}

\paragraph{Comparing networks} Another common question is if one can compare two different groups of people (e.g., clinical patients and healthy controls) regarding the connectivity or density of their networks (i.e. the number of edges)? The answer depends on the differences in sample size. As mentioned before, the EBIC is a function of the sample size: the lower the sample size, the more parsimonious the network structure. This means that comparing the connectivity of two networks is meaningful if they were estimated on roughly the same sample size, but that differences should not be compared if this assumption is not met (e.g., see \citealt{rhemtulla2016network}). One option is to perform a \emph{permutation test} \citep{NCT}. A permutation test is a a data-driven method in which all data are first pooled and then randomly assigned to two groups, resulting in two estimated networks. Repeating this process a number of times (e.g., 1000) leads to a distribution of differences between networks given that the two groups come from the same population. This distribution can subsequently be used to perform statistical tests on differences of the network structure between the two groups. The permutation test is implemented in the R package \emph{NetworkComparisonTest}. 

\paragraph{Network models versus latent variables} A final common question relates to how much network modeling and latent variable modeling overlap. Network modeling has been proposed as an alternative to latent variable modeling. As such, researchers might wonder if fitting a network model can provide evidence that the data are indeed generated by a system of variables causally influencing each-other, and not from a common cause model where the covariance between variables is explained by one or more underlying latent variables \citep{schmittmann2013}? The short answer is no. While psychological networks have been introduced as an alternative modeling framework to latent variable modeling, and are capable of strongly changing the point of focus from the common shared variance to unique variance between variables \citep{costantini2015state}, they cannot disprove the latent variable model. This is because there is a direct equivalence between network models and latent variable models \citep{epskampPsychometrika,netpsych,van2006dynamical,kruis2,marsman2015bayesian}. As discussed above, a latent variable causing covariation on multiple items should lead to a fully connected cluster of items if they are modeled as a network.

While the presence of a latent variable results in a fully connected cluster in the network, this does not mean that when the \emph{estimated} network does not contain fully connected clusters, the latent variable model must be false. As explained above, the LASSO retaining an edge can provide evidence that an edge is present, but not retaining an edge does not provide evidence that the edge is not present because an edge could simply not be estimated due to a lack of power. We refer the reader to \citet{boschlooCommentary} for a more detailed discussion on this topic and to \citet{epskampPsychometrika} for methodology on statistically comparing the fit of a network model and to that of a latent variable model.  Finally, just because two models are equivalent does not mean that they are equally plausible. For example, a lattice shaped network structure (nodes ordered on a grid and connected only to neighbors) is equivalent to some latent variable model, but the latent variable model is complicated and very implausible (many latent variables would be needed to explain the data; \citealt{marsman2015bayesian}).

Even when one expects a network model to largely explain the data, it may be implausible to assume that no latent variables cause any covariation in the network model \citep{friedchallenge,epskampPsychometrika,chandrasekaran2010latent}. To this end, estimating causal networks can lead to faulty causal hypotheses in the presence of latent variables. This issue is less problematic when estimating (undirected) partial correlation networks, as no direction of effect is coupled to the estimated edges. Methodologies to combine latent variable modeling and network modeling are currently being developed, which would allow researchers to use strengths from one framework to overcome weaknesses of the other framework. To overcome induced edges from latent variables, one can estimate a network structure after taking covariation due to one or more common causes into account (termed a \emph{residual} network; \citealt{panPsychMethods,epskampPsychometrika,fusedlasso,chandrasekaran2010latent}). Another way of combining network models with latent variable models is to use latent variables as nodes in a network (termed a \emph{latent} network; \citealt{epskampPsychometrika}). Doing so can cope with potential measurement error in the observed variables, allowing for powerful exploratory model search on the structural effects between latent variables \citep{guyon2017modeling}. Finally, statistical tests to distinguish sparse networks from latent variable models are currently being developed \citep{riettalk}.

% We would expect observed items caused by latent variables to result in clusters of fully connected nodes \citep{EGA,boschlooCommentary}. This demonstrates another use for partial correlation networks: in addition to discovering strong unique associations between observed variables, the network can also be indicative of latent variables by investigating the clustering in the network. 

\section{Conclusion}

This paper contains a tutorial on how to estimate psychological networks using a popular estimation technique: LASSO regularization with the EBIC model selection. This method provides a network of partial correlation coefficients with a limited number of spurious edges and can be based on either continuous or ordered-categorical data. This methodology has grown prominent in the past years and is featured in an increasing number of publications throughout various fields of psychological research. In addition, this paper (a) discusses in detail what partial correlations and partial correlation networks are and how these should be interpreted, (b) shows how researchers can estimate these network models in psychological datasets, (c) introduces a new simulation tool to perform power analysis for psychological networks, (d) summarizes post-hoc stability and accuracy analyses, and (e) describes how to deal with most commonly encountered issues when estimating and interpreting regularized partial correlation networks.

	The methods described in this paper are only appropriate when the cases in the data can reasonably be assumed to be independent. As this is plausible in cross-sectional analysis, we have exemplified the methodology by analyzing such a dataset. Several authors criticize cross-sectional analysis for not being able to separate within- and between-person variation \citep{molenaar2004manifesto,hamaker2012researchers,Bos}, and propose to study longitudinal data in order to capture within-person relationships \citep{bringmann2013}. We refer the reader to \citet{dynamics} for discussion on this topic and simulation studies studying cross-sectional analysis, and to \citet{weinberger} for a discussion on the causal interpretation of relationships when within-subject variation is lacking. The methods discussed in this paper can readily be applied to within-person data to obtain network structures not confounded by between-subject effects \citep{dynamics}. For a recent tutorial on this methodology, we refer the reader to \citet{costantini2017}. A downside of this method is that temporal information is not taken into account when estimating network structures. One way to estimate partial correlation networks while taking temporal information into account is by using the \emph{graphical vector-autoregression} model (graphical VAR; \citealt{wild2010graphical, Fisher2017, dynamics}), for which LASSO regularization techniques have been worked out \citep{abegaz2013sparse,rothman2010sparse}. EBIC model selection using these routines has been implemented in the R packages \emph{sparseTSCGM} (\citealt{SparseTSCGM}; aimed at estimating genetic networks) and \emph{graphicalVAR} (\citealt{graphicalVAR}; aimed at estimating $n=1$ psychological networks). 
	
The use of network modeling in psychology is still a young field and is not without challenges. Several related topics were beyond the scope of this tutorial and are discussed elsewhere in the literature. For an overview of challenges and future directions in network modeling of psychological data we refer the reader to \citet{friedchallenge} and \citet{roadAhead}. Psychological network analysis is a novel field that is rapidly changing and developing. We have not seen an accessible description of the most commonly used estimation procedure in the literature: LASSO regularization using EBIC model selection to estimate a sparse partial correlation network. This paper addresses this gap by providing an overview of this common and promising method.

\section*{Acknowledgements}

We would like to thank Jamie DeCoster for detailed feedback on earlier versions of this paper.

\bibliographystyle{apalike}
\bibliography{Bibliography}

\begin{thebibliography}{}

\bibitem[Abegaz and Wit, 2013]{abegaz2013sparse}
Abegaz, F. and Wit, E. (2013).
\newblock Sparse time series chain graphical models for reconstructing genetic
  networks.
\newblock {\em Biostatistics}, page kxt005.

\bibitem[Abegaz and Wit, 2015]{SparseTSCGM}
Abegaz, F. and Wit, E. (2015).
\newblock {\em {SparseTSCGM}: Sparse Time Series Chain Graphical Models}.
\newblock R package version 2.2.

\bibitem[Agresti, 1990]{agresti2014categorical}
Agresti, A. (1990).
\newblock {\em Categorical data analysis}.
\newblock John Wiley \& Sons Inc, New York, NY.

\bibitem[Armour et~al., 2017]{armour2017network}
Armour, C., Fried, E.~I., Deserno, M.~K., Tsai, J., and Pietrzak, R.~H. (2017).
\newblock A network analysis of dsm-5 posttraumatic stress disorder symptoms
  and correlates in us military veterans.
\newblock {\em Journal of anxiety disorders}, 45:49--59.

\bibitem[Bastian et~al., 2009]{bastian2009gephi}
Bastian, M., Heymann, S., Jacomy, M., et~al. (2009).
\newblock Gephi: an open source software for exploring and manipulating
  networks.
\newblock {\em Icwsm}, 8:361--362.

\bibitem[Borsboom, 2017]{borsboom2017network}
Borsboom, D. (2017).
\newblock A network theory of mental disorders.
\newblock {\em World Psychiatry}, 16(1):5--13.

\bibitem[Borsboom and Cramer, 2013]{borsboom2013network}
Borsboom, D. and Cramer, A. O.~J. (2013).
\newblock Network analysis: an integrative approach to the structure of
  psychopathology.
\newblock {\em Annual review of clinical psychology}, 9:91--121.

\bibitem[Borsboom et~al., in press]{forbescommentary}
Borsboom, D., Fried, E., Epskamp, S., Waldorp, L., van Borkulo, C., van~der
  Maas, H., and Cramer, A. O.~J. ({in press}).
\newblock Replicability of psychopathology networks: the right question but the
  wrong answer. a comment on ``evidence that psychopathology symptom networks
  have limited replicability'' by forbes, wright, markon, and krueger.
\newblock {\em Journal of Abnormal Psychology}.

\bibitem[Bos et~al., 2017]{Bos}
Bos, F.~M., Snippe, E., de~Vos, S., Hartmann, J.~A., Simons, C. J.~P., van~der
  Krieke, L., de~Jonge, P., and Wichers, M. (2017).
\newblock {Exploring the Idiographic Dynamics of Mood and Anxiety via Network
  Analysis}.

\bibitem[Bringmann et~al., 2013]{bringmann2013}
Bringmann, L.~F., Vissers, N., Wichers, M., Geschwind, N., Kuppens, P.,
  Peeters, F., Borsboom, D., and Tuerlinckx, F. (2013).
\newblock A network approach to psychopathology: New insights into clinical
  longitudinal data.
\newblock {\em PloS one}, 8(4):e60188.

\bibitem[Chandrasekaran et~al., 2012]{chandrasekaran2010latent}
Chandrasekaran, V., Parrilo, P.~A., and Willsky, A.~S. (2012).
\newblock Latent variable graphical model selection via convex optimization
  (with discussion).
\newblock {\em The Annals of Statistics}, 40(4):1935--1967.

\bibitem[Chapman et~al., 2016]{chapman2016statistical}
Chapman, B.~P., Weiss, A., and Duberstein, P.~R. (2016).
\newblock Statistical learning theory for high dimensional prediction:
  Application to criterion-keyed scale development.
\newblock {\em Psychological methods}, 21(4):603--620.

\bibitem[Chen and Chen, 2008]{chen2008EBIC}
Chen, J. and Chen, Z. (2008).
\newblock Extended bayesian information criteria for model selection with large
  model spaces.
\newblock {\em Biometrika}, 95(3):759--771.

\bibitem[Chen et~al., 2016]{fusedlasso}
Chen, Y., Li, X., Liu, J., and Ying, Z. (2016).
\newblock A fused latent and graphical model for multivariate binary data.
\newblock {\em arXiv preprint}, page arXiv:1606.08925.

\bibitem[Chernick, 2011]{chernick2011bootstrap}
Chernick, M.~R. (2011).
\newblock {\em Bootstrap methods: A guide for practitioners and researchers},
  volume 619.
\newblock John Wiley \& Sons, New York, NY, USA.

\bibitem[Cohen, 1977]{cohen1977statistical}
Cohen, J. (1977).
\newblock {\em Statistical power analysis for the behavioral sciences}.
\newblock Academic Press, New York, NY, USA.

\bibitem[Cohen et~al., 2003]{cohen2013applied}
Cohen, J., Cohen, P., West, S.~G., and Aiken, L.~S. (2003).
\newblock {\em Applied multiple regression/correlation analysis for the
  behavioral sciences}.
\newblock Erlbaum, Hillsdale, NJ.

\bibitem[Costantini et~al., 2015a]{costantini2015state}
Costantini, G., Epskamp, S., Borsboom, D., Perugini, M., M\~{o}ttus, R.,
  Waldorp, L.~J., and Cramer, A. O.~J. (2015a).
\newblock State of the {aRt} personality research: A tutorial on network
  analysis of personality data in {R}.
\newblock {\em Journal of Research in Personality}, 54:13--29.

\bibitem[Costantini et~al., 2015b]{costantini2015development}
Costantini, G., Richetin, J., Borsboom, D., Fried, E.~I., Rhemtulla, M., and
  Perugini, M. (2015b).
\newblock Development of indirect measures of conscientiousness: combining a
  facets approach and network analysis.
\newblock {\em European Journal of Personality}, 29(5):548--567.

\bibitem[Costantini et~al., 2017]{costantini2017}
Costantini, G., Richetin, J., Preti, E., Casini, E., Epskamp, S., and Perugini,
  M. (2017).
\newblock Stability and variability of personality networks. a tutorial on
  recent developments in network psychometrics.
\newblock {\em Personality and Individual Differences}.

\bibitem[Cox and Wermuth, 1994]{cox1994note}
Cox, D.~R. and Wermuth, N. (1994).
\newblock A note on the quadratic exponential binary distribution.
\newblock {\em Biometrika}, 81(2):403--408.

\bibitem[Cramer et~al., 2012]{cramer2012dimensions}
Cramer, A. O.~J., Sluis, S., Noordhof, A., Wichers, M., Geschwind, N., Aggen,
  S.~H., Kendler, K.~S., and Borsboom, D. (2012).
\newblock Dimensions of normal personality as networks in search of
  equilibrium: You can't like parties if you don't like people.
\newblock {\em European Journal of Personality}, 26(4):414--431.

\bibitem[Cramer et~al., 2010]{cramer2010comorbidity}
Cramer, A. O.~J., Waldorp, L., van~der Maas, H., and Borsboom, D. (2010).
\newblock {Comorbidity: A Network Perspective}.
\newblock {\em Behavioral and Brain Sciences}, 33(2-3):137--150.

\bibitem[Csardi and Nepusz, 2006]{igraph}
Csardi, G. and Nepusz, T. (2006).
\newblock The {igraph} software package for complex network research.
\newblock {\em InterJournal}, Complex Systems:1695.

\bibitem[Deserno et~al., 2016]{Deserno17082016}
Deserno, M.~K., Borsboom, D., Begeer, S., and Geurts, H.~M. (2016).
\newblock Multicausal systems ask for multicausal approaches: A network
  perspective on subjective well-being in individuals with autism spectrum
  disorder.
\newblock {\em Autism}.

\bibitem[Drton and Perlman, 2004]{drton2004model}
Drton, M. and Perlman, M.~D. (2004).
\newblock Model selection for gaussian concentration graphs.
\newblock {\em Biometrika}, 91(3):591--602.

\bibitem[Dziak et~al., 2012]{dziak2012sensitivity}
Dziak, J.~J., Coffman, D.~L., Lanza, S.~T., and Li, R. (2012).
\newblock Sensitivity and specificity of information criteria.
\newblock {\em The Methodology Center and Department of Statistics, Penn State,
  The Pennsylvania State University}.

\bibitem[Efron, 1979]{efron1979bootstrap}
Efron, B. (1979).
\newblock Bootstrap methods: another look at the jackknife.
\newblock {\em The Annals of Statistics}, 7(1):1--26.

\bibitem[Epskamp, 2015]{graphicalVAR}
Epskamp, S. (2015).
\newblock {\em {graphicalVAR}: Graphical VAR for Experience Sampling Data}.
\newblock R package version 0.1.3.

\bibitem[Epskamp, 2016]{qgraphsims}
Epskamp, S. (2016).
\newblock {Regularized Gaussian Psychological Networks: Brief Report on the
  Performance of Extended BIC Model Selection}.
\newblock {\em arXiv preprint}, page arXiv:1606.05771.

\bibitem[Epskamp, 2017]{roadAhead}
Epskamp, S. (2017).
\newblock Discussion: The road ahead.
\newblock In Epskamp, S., editor, {\em Network Psychometrics}, chapter~12. PhD
  Dissertation.

\bibitem[Epskamp et~al., 2017a]{bootnetpaper}
Epskamp, S., Borsboom, D., and Fried, E.~I. (2017a).
\newblock {Estimating psychological networks and their accuracy: a tutorial
  paper}.
\newblock {\em Behavior Research Methods}.

\bibitem[Epskamp et~al., 2012]{jssv048i04}
Epskamp, S., Cramer, A., Waldorp, L., Schmittmann, V.~D., and Borsboom, D.
  (2012).
\newblock qgraph: Network visualizations of relationships in psychometric data.
\newblock {\em Journal of Statistical Software}, 48(1):1--18.

\bibitem[Epskamp et~al., 2017b]{boschlooCommentary}
Epskamp, S., Kruis, J., and Marsman, M. (2017b).
\newblock Estimating psychopathological networks: be careful what you wish for.
\newblock {\em PlosOne}, 12(6):e0179891.

\bibitem[Epskamp et~al., 2016]{netpsych}
Epskamp, S., Maris, G., Waldorp, L., and Borsboom, D. (2016).
\newblock Network psychometrics.
\newblock In Irwing, P., Hughes, D., and Booth, T., editors, {\em Handbook of
  Psychometrics}. Wiley, New York, NY, USA.

\bibitem[Epskamp et~al., 2017c]{epskampPsychometrika}
Epskamp, S., Rhemtulla, M., and Borsboom, D. (2017c).
\newblock Generalized network psychometrics: Combining network and latent
  variable models.
\newblock {\em Psychometrika}.

\bibitem[Epskamp et~al., 2017d]{dynamics}
Epskamp, S., Waldorp, L.~J., M\~{o}ttus, R., and Borsboom, D. (2017d).
\newblock Discovering psychological dynamics in time-series data.
\newblock {\em arXiv preprint}, page arXiv:1609.04156.

\bibitem[Fisher et~al., 2017]{Fisher2017}
Fisher, A.~J., Reeves, J.~W., Lawyer, G., Medaglia, J.~D., and Rubel, J.~A.
  (2017).
\newblock {Exploring the Idiographic Dynamics of Mood and Anxiety via Network
  Analysis}.

\bibitem[Forbes et~al., in press]{forbes}
Forbes, M.~K., Wright, A. G.~C., Markon, K., and Krueger, R. (in press).
\newblock Evidence that psychopathology symptom networks have limited
  replicability.
\newblock {\em Journal of Abnormal Psychology}.

\bibitem[Foygel and Drton, 2010]{foygel2010extended}
Foygel, R. and Drton, M. (2010).
\newblock Extended {Bayesian} information criteria for {Gaussian} graphical
  models.
\newblock {\em Advances in Neural Information Processing Systems},
  23:2020--2028.

\bibitem[Foygel~Barber and Drton, 2015]{foygel2014high}
Foygel~Barber, R. and Drton, M. (2015).
\newblock High-dimensional {Ising} model selection with bayesian information
  criteria.
\newblock {\em Electronic Journal of Statistics}, 9(1):567–607.

\bibitem[Fried and Cramer, 2017]{friedchallenge}
Fried, E.~I. and Cramer, A. O.~J. (2017).
\newblock Moving forward: challenges and directions for psychopathological
  network theory and methodology.
\newblock {\em Perspectives on Psychological Science}.

\bibitem[Fried et~al., 2017a]{fried2017replicability}
Fried, E.~I., Eidhof, M.~B., Palic, S., Costantini, G., Huisman-van Diujk,
  H.~M., Bockting, C. L.~M., Engelhard, I., Armour, C., Nielsen, A. B.~S., and
  Karstoft, K.-I. (2017a).
\newblock Replicability and generalizability of ptsd networks: A cross-cultural
  multisite study of ptsd symptoms in four trauma patient samples.

\bibitem[Fried et~al., 2016]{fried2016}
Fried, E.~I., Epskamp, S., Nesse, R.~M., Tuerlinckx, F., and Borsboom, D.
  (2016).
\newblock {What are `good' depression symptoms? Comparing the centrality of DSM
  and non-DSM symptoms of depression in a network analysis}.
\newblock {\em Journal of Affective Disorders}, 189:314--320.

\bibitem[Fried et~al., 2017b]{friedreview}
Fried, E.~I., van Borkulo, C.~D., Cramer, A. O.~J., Lynn, B., Schoevers, R.~A.,
  and Borsboom, D. (2017b).
\newblock Mental disorders as networks of problems: a review of recent
  insights.
\newblock {\em Social Psychiatry and Psychiatric Epidemiology}, 52:1--10.

\bibitem[Friedman et~al., 2008]{friedman2008sparse}
Friedman, J.~H., Hastie, T., and Tibshirani, R. (2008).
\newblock Sparse inverse covariance estimation with the graphical lasso.
\newblock {\em Biostatistics}, 9(3):432--441.

\bibitem[Friedman et~al., 2014]{glasso}
Friedman, J.~H., Hastie, T., and Tibshirani, R. (2014).
\newblock {\em glasso: Graphical lasso- estimation of Gaussian graphical
  models}.
\newblock R package version 1.8.

\bibitem[Fruchterman and Reingold, 1991]{fruchterman1991graph}
Fruchterman, T. and Reingold, E. (1991).
\newblock Graph drawing by force-directed placement.
\newblock {\em Software: Practice and Experience}, 21(11):1129--1164.

\bibitem[Gentry et~al., 2011]{rgraphviz}
Gentry, J., Long, L., Gentleman, R., Falcon, S., Hahne, F., Sarkar, D., and
  Hansen, K. (2011).
\newblock {\em Rgraphviz: Provides Plotting Capabilities for R Graph Objects}.
\newblock R package version 1.27.0.

\bibitem[Golino and Epskamp, 2017]{golino2017exploratory}
Golino, H.~F. and Epskamp, S. (2017).
\newblock Exploratory graph analysis: A new approach for estimating the number
  of dimensions in psychological research.
\newblock {\em PloS one}, 12(6):e0174035.

\bibitem[Guyon et~al., 2017]{guyon2017modeling}
Guyon, H., Falissard, B., and Kop, J.-L. (2017).
\newblock Modeling psychological attributes in psychology--an epistemological
  discussion: Network analysis vs. latent variables.
\newblock {\em Frontiers in Psychology}, 8.

\bibitem[Hamaker, 2012]{hamaker2012researchers}
Hamaker, E.~L. (2012).
\newblock Why researchers should think “within-person”: A paradigmatic
  rationale.
\newblock {\em Handbook of research methods for studying daily life}, pages
  43--61.

\bibitem[Haslbeck and Waldorp, 2016]{mgm}
Haslbeck, J. M.~B. and Waldorp, L.~J. (2016).
\newblock {mgm}: Structure estimation for time-varying mixed graphical models
  in high-dimensional data.
\newblock {\em arXiv preprint}, page arXiv:1510.06871.

\bibitem[Hastie et~al., 2001]{hastie01statisticallearning}
Hastie, T., Tibshirani, R., and Friedman, J. (2001).
\newblock {\em {The Elements of Statistical Learning}}.
\newblock Springer Series in Statistics. Springer New York Inc., New York, NY,
  USA.

\bibitem[Hastie et~al., 2015]{hastie2015statistical}
Hastie, T., Tibshirani, R., and Wainwright, M. (2015).
\newblock {\em Statistical learning with sparsity: the lasso and
  generalizations}.
\newblock CRC Press, Boca Raton, FL, USA.

\bibitem[Holland and Rosenbaum, 1986]{holland1986conditional}
Holland, P.~W. and Rosenbaum, P.~R. (1986).
\newblock Conditional association and unidimensionality in monotone latent
  variable models.
\newblock {\em The Annals of Statistics}, 14:1523--1543.

\bibitem[Isvoranu et~al., 2016a]{isvoranua}
Isvoranu, A.~M., Borsboom, D., van Os, J., and Guloksuz, S. (2016a).
\newblock {A Network Approach to Environmental Impact in Psychotic Disorders:
  Brief Theoretical Framework.}
\newblock {\em Schizophrenia Bulletin}, 42(4):870--873.

\bibitem[Isvoranu et~al., 2016b]{isvoranu}
Isvoranu, A.~M., van Borkulo, C.~D., Boyette, L., Wigman, J. T.~W., Vinkers,
  C.~H., Borsboom, D., and {GROUP Investigators} (2016b).
\newblock {A Network Approach to Psychosis: Pathways between Childhood Trauma
  and Psychotic Symptoms}.
\newblock {\em Schizophrenia Bulletin}.
\newblock Advance Access published May 10, 2016.

\bibitem[Jaya et~al., 2016]{jaya2015loneliness}
Jaya, E.~S., Hillmann, T.~E., Reininger, K.~M., Gollwitzer, A., and Lincoln,
  T.~M. (2016).
\newblock Loneliness and psychotic symptoms: The mediating role of depression.
\newblock {\em Cognitive Therapy and Research}.

\bibitem[Knefel et~al., 2016]{knefel2016association}
Knefel, M., Tran, U.~S., and Lueger-Schuster, B. (2016).
\newblock The association of posttraumatic stress disorder, complex
  posttraumatic stress disorder, and borderline personality disorder from a
  network analytical perspective.
\newblock {\em Journal of Anxiety Disorders}, 43:70--78.

\bibitem[Koller and Friedman, 2009]{koller2009probabilistic}
Koller, D. and Friedman, N. (2009).
\newblock {\em Probabilistic graphical models: principles and techniques}.
\newblock MIT press, Cambridge, MA, USA.

\bibitem[Kossakowski et~al., 2015]{kossakowski2015}
Kossakowski, J.~J., Epskamp, S., Kieffer, J.~M., van Borkulo, C.~D., Rhemtulla,
  M., and Borsboom, D. (2015).
\newblock The application of a network approach to health-related quality of
  life ({HRQoL}): Introducing a new method for assessing hrqol in healthy
  adults and cancer patient.
\newblock {\em Quality of Life Research}, 25:781--92.

\bibitem[Kr{\"a}mer et~al., 2009]{parcor}
Kr{\"a}mer, N., Sch{\"a}fer, J., and Boulesteix, A.-L. (2009).
\newblock Regularized estimation of large-scale gene association networks using
  graphical gaussian models.
\newblock {\em BMC Bioinformatics}, 10(1):1--24.

\bibitem[Kruis and Maris, 2016]{kruis2}
Kruis, J. and Maris, G. (2016).
\newblock Three representations of the ising model.
\newblock {\em Scientific Reports}, 6.

\bibitem[Langley et~al., 2015]{langley2015should}
Langley, D.~J., Wijn, R., Epskamp, S., and {Van Bork}, R. (2015).
\newblock {Should I Get That Jab? Exploring Influence to Encourage Vaccination
  via Online Social Media}.
\newblock {\em ECIS 2015 Research-in-Progress Papers}, page Paper 64.

\bibitem[Lauritzen, 1996]{lauritzen1996graphical}
Lauritzen, S.~L. (1996).
\newblock {\em Graphical models}.
\newblock Clarendon Press, Oxford, UK.

\bibitem[Levine and Leucht, 2016]{levine2016identifying}
Levine, S.~Z. and Leucht, S. (2016).
\newblock Identifying a system of predominant negative symptoms: Network
  analysis of three randomized clinical trials.
\newblock {\em Schizophrenia Research}.

\bibitem[Liu et~al., 2009]{liu2009nonparanormal}
Liu, H., Lafferty, J.~D., and Wasserman, L. (2009).
\newblock The nonparanormal: Semiparametric estimation of high dimensional
  undirected graphs.
\newblock {\em The Journal of Machine Learning Research}, 10:2295--2328.

\bibitem[Marsman et~al., 2015]{marsman2015bayesian}
Marsman, M., Maris, G., Bechger, T., and Glas, C. (2015).
\newblock Bayesian inference for low-rank ising networks.
\newblock {\em Scientific reports}, 5(9050):1--7.

\bibitem[McNally, 2016]{mcnally2016can}
McNally, R.~J. (2016).
\newblock Can network analysis transform psychopathology?
\newblock {\em Behaviour Research and Therapy}.

\bibitem[McNally et~al., 2015]{mcnally2015mental}
McNally, R.~J., Robinaugh, D.~J., Wu, G.~W., Wang, L., Deserno, M.~K., and
  Borsboom, D. (2015).
\newblock Mental disorders as causal systems a network approach to
  posttraumatic stress disorder.
\newblock {\em Clinical Psychological Science}, 3(6):836--849.

\bibitem[Meinshausen and B{\"u}hlmann, 2006]{meinshausen2006high}
Meinshausen, N. and B{\"u}hlmann, P. (2006).
\newblock High-dimensional graphs and variable selection with the lasso.
\newblock {\em The annals of statistics}, pages 1436--1462.

\bibitem[Molenaar, 2004]{molenaar2004manifesto}
Molenaar, P.~C. (2004).
\newblock A manifesto on psychology as idiographic science: Bringing the person
  back into scientific psychology, this time forever.
\newblock {\em Measurement}, 2(4):201--218.

\bibitem[M{\~o}ttus and Allerhand, 2017]{mottus2017traits}
M{\~o}ttus, R. and Allerhand, M. (2017).
\newblock Why do traits come together? the underlying trait and network
  approaches.
\newblock {\em SAGE handbook of personality and individual differences},
  1:1--22.

\bibitem[Murphy, 2012]{murphy2012machine}
Murphy, K.~P. (2012).
\newblock {\em Machine learning: a probabilistic perspective}.
\newblock MIT press, Cambridge, MA, USA.

\bibitem[Muth{\'e}n, 1984]{muthen1984general}
Muth{\'e}n, B. (1984).
\newblock A general structural equation model with dichotomous, ordered
  categorical, and continuous latent variable indicators.
\newblock {\em Psychometrika}, 49(1):115--132.

\bibitem[Muth{\'e}n, 1989]{muthen1989factor}
Muth{\'e}n, B.~O. (1989).
\newblock Factor structure in groups selected on observed scores.
\newblock {\em British journal of mathematical and statistical psychology},
  42(1):81--90.

\bibitem[Olsson, 1979]{olsson1979maximum}
Olsson, U. (1979).
\newblock Maximum likelihood estimation of the polychoric correlation
  coefficient.
\newblock {\em Psychometrika}, 44(4):443--460.

\bibitem[Olsson et~al., 1982]{olsson1982polyserial}
Olsson, U., Drasgow, F., and Dorans, N.~J. (1982).
\newblock The polyserial correlation coefficient.
\newblock {\em Psychometrika}, 47(3):337--347.

\bibitem[{Open Science Collaboration}, 2015]{opensciencecollaboration2015}
{Open Science Collaboration} (2015).
\newblock {Estimating the reproducibility of psychological science}.
\newblock {\em Science}, 349(6251):aac4716--aac4716.

\bibitem[Opsahl et~al., 2010]{opsahl2010node}
Opsahl, T., Agneessens, F., and Skvoretz, J. (2010).
\newblock Node centrality in weighted networks: Generalizing degree and
  shortest paths.
\newblock {\em Social Networks}, 32(3):245--251.

\bibitem[Pan et~al., in press]{panPsychMethods}
Pan, J., Ip, E., and Dube, L. (In press).
\newblock An alternative to post-hoc model modification in confirmatory factor
  analysis: The bayesian lasso.
\newblock {\em Psychological Methods}.

\bibitem[Pearl, 2000]{pearl2000causality}
Pearl, J. (2000).
\newblock {\em {Causality: Models, Reasoning, and Inference}}.
\newblock Cambridge Univ Pr.

\bibitem[Pedersen, 2017]{ggraph}
Pedersen, T.~L. (2017).
\newblock {\em ggraph: An Implementation of Grammar of Graphics for Graphs and
  Networks}.
\newblock R package version 1.0.0.

\bibitem[Pourahmadi, 2011]{pourahmadi2011covariance}
Pourahmadi, M. (2011).
\newblock Covariance estimation: The glm and regularization perspectives.
\newblock {\em Statistical Science}, 26(3):369--387.

\bibitem[{R Core Team}, 2016]{R}
{R Core Team} (2016).
\newblock {\em R: A Language and Environment for Statistical Computing}.
\newblock R Foundation for Statistical Computing, Vienna, Austria.

\bibitem[Rhemtulla et~al., 2016]{rhemtulla2016network}
Rhemtulla, M., Fried, E.~I., Aggen, S.~H., Tuerlinckx, F., Kendler, K.~S., and
  Borsboom, D. (2016).
\newblock {Network analysis of substance abuse and dependence symptoms}.
\newblock {\em Drug and alcohol dependence}, 161:230--237.

\bibitem[Rigdon and Ferguson~Jr, 1991]{rigdon1991performance}
Rigdon, E.~E. and Ferguson~Jr, C.~E. (1991).
\newblock The performance of the polychoric correlation coefficient and
  selected fitting functions in confirmatory factor analysis with ordinal data.
\newblock {\em Journal of marketing research}, pages 491--497.

\bibitem[Rosseel, 2012]{lavaan}
Rosseel, Y. (2012).
\newblock {lavaan}: An {R} package for structural equation modeling.
\newblock {\em Journal of Statistical Software}, 48(2):1--36.

\bibitem[Rothman et~al., 2010]{rothman2010sparse}
Rothman, A.~J., Levina, E., and Zhu, J. (2010).
\newblock Sparse multivariate regression with covariance estimation.
\newblock {\em Journal of Computational and Graphical Statistics},
  19(4):947--962.

\bibitem[Schmittmann et~al., 2013]{schmittmann2013}
Schmittmann, V.~D., Cramer, A. O.~J., Waldorp, L.~J., Epskamp, S., Kievit,
  R.~A., and Borsboom, D. (2013).
\newblock {Deconstructing the construct: A network perspective on psychological
  phenomena}.
\newblock {\em New Ideas in Psychology}, 31(1):43--53.

\bibitem[Shannon et~al., 2003]{shannon2003cytoscape}
Shannon, P., Markiel, A., Ozier, O., Baliga, N.~S., Wang, J.~T., Ramage, D.,
  Amin, N., Schwikowski, B., and Ideker, T. (2003).
\newblock Cytoscape: a software environment for integrated models of
  biomolecular interaction networks.
\newblock {\em Genome research}, 13(11):2498--2504.

\bibitem[Stevens, 1946]{stevens1946theory}
Stevens, S.~S. (1946).
\newblock On the theory of scales of measurement.
\newblock {\em Science, New Series}, 103(2684):677--680.

\bibitem[Tibshirani, 1996]{tibshirani1996regression}
Tibshirani, R. (1996).
\newblock Regression shrinkage and selection via the lasso.
\newblock {\em Journal of the Royal Statistical Society. Series B
  (Methodological)}, 58:267--288.

\bibitem[Van~Bork, 2015]{riettalk}
Van~Bork, R. (2015).
\newblock Latent variable and network model implications for partial
  correlation structures.
\newblock In {\em 80th Annual Meeting of the {Psychometric Society (IMPS)}}.

\bibitem[Van~Borkulo et~al., 2017]{NCT}
Van~Borkulo, C., Boschloo, L., Kossakowski, J., Tio, P., Schoevers, R.,
  Borsboom, D., and Waldorp, L. (2017).
\newblock Comparing network structures on three aspects: A permutation test.

\bibitem[van Borkulo et~al., 2014]{van2014new}
van Borkulo, C.~D., Borsboom, D., Epskamp, S., Blanken, T.~F., Boschloo, L.,
  Schoevers, R.~A., and Waldorp, L.~J. (2014).
\newblock A new method for constructing networks from binary data.
\newblock {\em Scientific reports}, 4(5918):1--10.

\bibitem[van Borkulo et~al., 2015]{van2015association}
van Borkulo, C.~D., Boschloo, L., Borsboom, D., Penninx, B. W. J.~H., Waldorp,
  L.~J., and Schoevers, R.~A. (2015).
\newblock {Association of Symptom Network Structure With the Course of
  Depression}.
\newblock {\em JAMA psychiatry}, 72(12):1219--1226.

\bibitem[van Borkulo and Epskamp, 2014]{IsingFit}
van Borkulo, C.~D. and Epskamp, S. (2014).
\newblock {\em {IsingFit}: Fitting {Ising} models using the eLasso method}.
\newblock R package version 0.2.0.

\bibitem[Van Der~Maas et~al., 2017]{van2017network}
Van Der~Maas, H., Kan, K.-J., Marsman, M., and Stevenson, C.~E. (2017).
\newblock Network models for cognitive development and intelligence.

\bibitem[Van Der~Maas et~al., 2006]{van2006dynamical}
Van Der~Maas, H.~L., Dolan, C.~V., Grasman, R.~P., Wicherts, J.~M., Huizenga,
  H.~M., and Raijmakers, M.~E. (2006).
\newblock A dynamical model of general intelligence: the positive manifold of
  intelligence by mutualism.
\newblock {\em Psychological review}, 113(4):842--861.

\bibitem[Wagenmakers, 2007]{wagenmakers2007practical}
Wagenmakers, E.-J. (2007).
\newblock A practical solution to the pervasive problems ofp values.
\newblock {\em Psychonomic bulletin \& review}, 14(5):779--804.

\bibitem[Wasserman and Faust, 1994]{wasserman1994}
Wasserman, S. and Faust, K. (1994).
\newblock {\em {Social network analysis: Methods and applications}}.
\newblock Cambridge University Press, Cambridge, UK.

\bibitem[Weinberger, 2015]{weinberger}
Weinberger, N. (2015).
\newblock If intelligence is a cause, it is a within-subjects cause.
\newblock {\em Theory \& Psychology}, 25(3):346--361.

\bibitem[Wetzels and Wagenmakers, 2012]{wetzels2012default}
Wetzels, R. and Wagenmakers, E.-J. (2012).
\newblock A default bayesian hypothesis test for correlations and partial
  correlations.
\newblock {\em Psychonomic bulletin \& review}, 19(6):1057--1064.

\bibitem[Wickens, 1989]{wickens2014multiway}
Wickens, T.~D. (1989).
\newblock {\em Multiway contingency tables analysis for the social sciences}.
\newblock Lawrence Erlbaum Associates, Hillsdale, NJ, USA.

\bibitem[Wild et~al., 2010]{wild2010graphical}
Wild, B., Eichler, M., Friederich, H.-C., Hartmann, M., Zipfel, S., and Herzog,
  W. (2010).
\newblock {A graphical vector autoregressive modelling approach to the analysis
  of electronic diary data}.
\newblock {\em BMC medical research methodology}, 10(1):28.

\bibitem[Zhao and Yu, 2006]{zhao2006model}
Zhao, P. and Yu, B. (2006).
\newblock On model selection consistency of lasso.
\newblock {\em The Journal of Machine Learning Research}, 7:2541--2563.

\bibitem[Zhao et~al., 2015]{huge}
Zhao, T., Li, X., Liu, H., Roeder, K., Lafferty, J., and Wasserman, L. (2015).
\newblock {\em huge: High-Dimensional Undirected Graph Estimation}.
\newblock R package version 1.2.7.

\bibitem[Zou, 2006]{zou2006adaptive}
Zou, H. (2006).
\newblock The adaptive lasso and its oracle properties.
\newblock {\em Journal of the American statistical association},
  101(476):1418--1429.

\end{thebibliography}

\end{document}